\UseRawInputEncoding
\documentclass[aps,prl,floatfix,twocolumn,groupedaddress]{revtex4-1}
\usepackage{graphicx}
\usepackage{epstopdf}
\usepackage{times}
\usepackage{amssymb,amsmath}
\usepackage{array}
\usepackage{multirow}
\usepackage[flushleft]{caption}
\usepackage{color}
\usepackage{setspace}

\makeatletter

\newcommand{\ket}[1]{\left\vert#1\right\rangle}

\begin{document}%

\title{Three-dimensional entanglement on a silicon chip}

\author{Liangliang Lu$^{1,\dagger}$, Lijun Xia$^{1,\dagger}$, Zhiyu Chen$^{1,\dagger}$, Leizhen Chen$^{1}$, Tonghua Yu$^{1}$, Tao Tao$^{1}$, Wenchao Ma$^{1}$, Ying Pan$^{2}$, Xinlun Cai$^{2}$, Yanqing Lu$^{1}$, Shining Zhu$^{1}$, Xiao-Song Ma$^{1,\ast}$}
\affiliation{
$^1$~National Laboratory of Solid-state Microstructures, School of Physics, College of Engineering and Applied Sciences, Collaborative Innovation Center of Advanced Microstructures, Nanjing University, Nanjing 210093, China\\
$^2$~State Key Laboratory of Optoelectronic Materials and Technologies and School of Physics and Engineering, Sun Yat-sen University, Guangzhou 510275, China\\
$^{\dagger}$These authors contributed equally to this work\\
$^{\ast}$e-mail:Xiaosong.Ma@nju.edu.cn}


\begin{abstract}	

Entanglement is a counterintuitive feature of quantum physics that is at the heart of quantum technology. High-dimensional quantum states offer unique advantages in various quantum information tasks. Integrated photonic chips have recently emerged as a leading platform for the generation, manipulation and detection of entangled photons. Here, we report a silicon photonic chip that uses interferometric resonance-enhanced photon-pair sources, spectral demultiplexers and high-dimensional reconfigurable circuitries to generate, manipulate and analyse path-entangled three-dimensional qutrit states. By minimizing on-chip electrical and thermal cross-talk, we obtain high-quality quantum interference with visibilities above 96.5\% and a maximally entangled qutrit state with a fidelity of 95.5\%. We further explore the fundamental properties of entangled qutrits to test quantum nonlocality and contextuality, and to implement quantum simulations of graphs and high-precision optical phase measurements. Our work paves the path for the development of multiphoton high-dimensional quantum technologies.

\end{abstract}

\maketitle
\onecolumngrid

\section*{Introduction}

Entanglement is a central resource for quantum-enhanced technology, including quantum computation~\cite{ladd2010}, communication~\cite{gisin2002} and metrology~\cite{giovannetti2011}. To demonstrate the advantage of quantum systems, it is necessary to generate, manipulate and detect entangled states. Quantum states span the Hilbert space with a dimensionality of d${}^{n}$, where $d$ is the dimensionality of a single particle and n is the number of particles in the entangled states. Most of the widely-used quantum information processing protocols are based on qubits, a quantum system with d=2. Recently, higher-dimensional entangled states (qudits, d$\mathrm{>}$2) have gained substantial interest, owing to their distinguishing properties. For example, qudits provide larger channel capacity and better noise tolerance in quantum communication~\cite{dambrosio2012, graham2015,luo2019,hu2019experimental}, as well as higher efficiency and flexibility in quantum computing~\cite{lanyon2009, qiang2018} and simulations~\cite{neeley2009}. From the fundamental point of view, qudits also provide stronger violations of Bell inequalities~\cite{dada2011}, lower bounds for closing the fair-sampling loopholes in Bell tests~\cite{vertesi2010} and possibilities to test contextuality~\cite{lapkiewicz2011}. Recent reviews on the high-dimensional entanglement can be found in refs.~\cite{erhard2019,wang2019integrated}.

High-dimensional entangled photons have been realized in various degrees of freedom (DOFs), including orbital angular momentum (OAM)~\cite{dada2011, wang2015}, frequency~\cite{kues2017,imany2018}, path~\cite{schaeff2015, wang2018}, temporal~\cite{Thew:2004:ERE:2011577.2011578, Richart:2012hk} and hybrid time-frequency modes~\cite{reimer2019}. In particular, path-entangled photon pairs have been studied with a view to quantum information processing, where they are particularly attractive due to their conceptual simplicity~\cite{Reck:1994wd}. However, the generation of high-dimensional path-entangled photon pairs typically requires the simultaneous operation of several coherently pumped indistinguishable photon-pair sources and several multi-path interferometers with high phase stability~\cite{schaeff2012}. As the dimensionality increases, the phase stabilization quickly becomes a daunting task in experiments based on bulk and fibre optical elements.

Integrated photonic circuits based on silicon offer dense component integration, high optical non-linearity and good phase stability, which are highly desirable properties for photonic quantum technology~\cite{mower2011,collins2013, Silverstone:2014fu,  Harris:2014wa, Silverstone:2015cl, feng2016}. Moreover, silicon photonic devices are routinely fabricated in complementary metal oxide semiconductor (CMOS) processes. Therefore, a new field called silicon quantum photonics has recently been developed and has emerged as a promising platform for large-scale quantum information processing~\cite{silverstone2016}. Recent advances of on-chip high-dimensional entanglement have employed frequency-encoding generated from a micro-resonator photon-pair source~\cite{kues2017} and path-encoding generated from meander waveguides photon-pair source~\cite{wang2018}. Specifically, silicon waveguides with cm-length are often employed as sources to create photon-pairs~\cite{qiang2018,wang2018,Silverstone:2014fu}. However, the natural bandwidth of the photons generated from meander waveguides is about 30 nm~\cite{wang2018}. In order to obtain high-quality photons, it is necessary to employ band-pass filters ($\sim$1nm bandwidth in ref.~\cite{wang2018}), which unavoidably reduces the photon count rate drastically.

In this work, we employ a silicon photonic chip using an advanced resonator source embedded in Mach-Zehnder interferometers (MZIs) to generate, manipulate and characterize path-entangled qutrits (d=3). Cavity-enhanced processes and independent tuning capabilities of the coupling coefficients of pump, signal and idler photons allow us to generate high-indistinguishable and high-brightness photons without using passive filtering. The bandwidth of the photon generated from our source is about 50 pm, about a factor of 600 narrower to that of ref.~\cite{wang2018}. This narrow-band feature not only provides high-quality single photons, but also holds the promise for direct coupling with telecom quantum memory~\cite{saglamyurek2015}, which is not possible for the nanowire source due to the prohibitive low count rate after $\sim$GHz bandwidth filtering. In particular, we perform on-chip test of quantum contextuality with closed compatible loophole. Furthermore, using the entangled qutrit state, we simulate a two-vertex and three-edge graph and obtain the number of the perfect matchings of this graph, which is in the $\#$P-complete complexity class~\cite{valiant1979}. Although the structure of the graph of our demonstration is simple, it can be viewed as the first step towards achieving the ambitious goal of solving $\#$P hard problem with quantum photonic devices. We also employ our device to demonstrate the excellent phase sensitivity, exceeding both classical three-path linear interferometer and quantum second-order nonlinear interferometer limits. To be best of our knowledge, none of these three experiments have ever been realized with an integrated chip.

\section*{Results}

\textbf{Silicon quantum photonic chip and experimental setup}
\begin{figure}
    \includegraphics[width=0.85\textwidth]{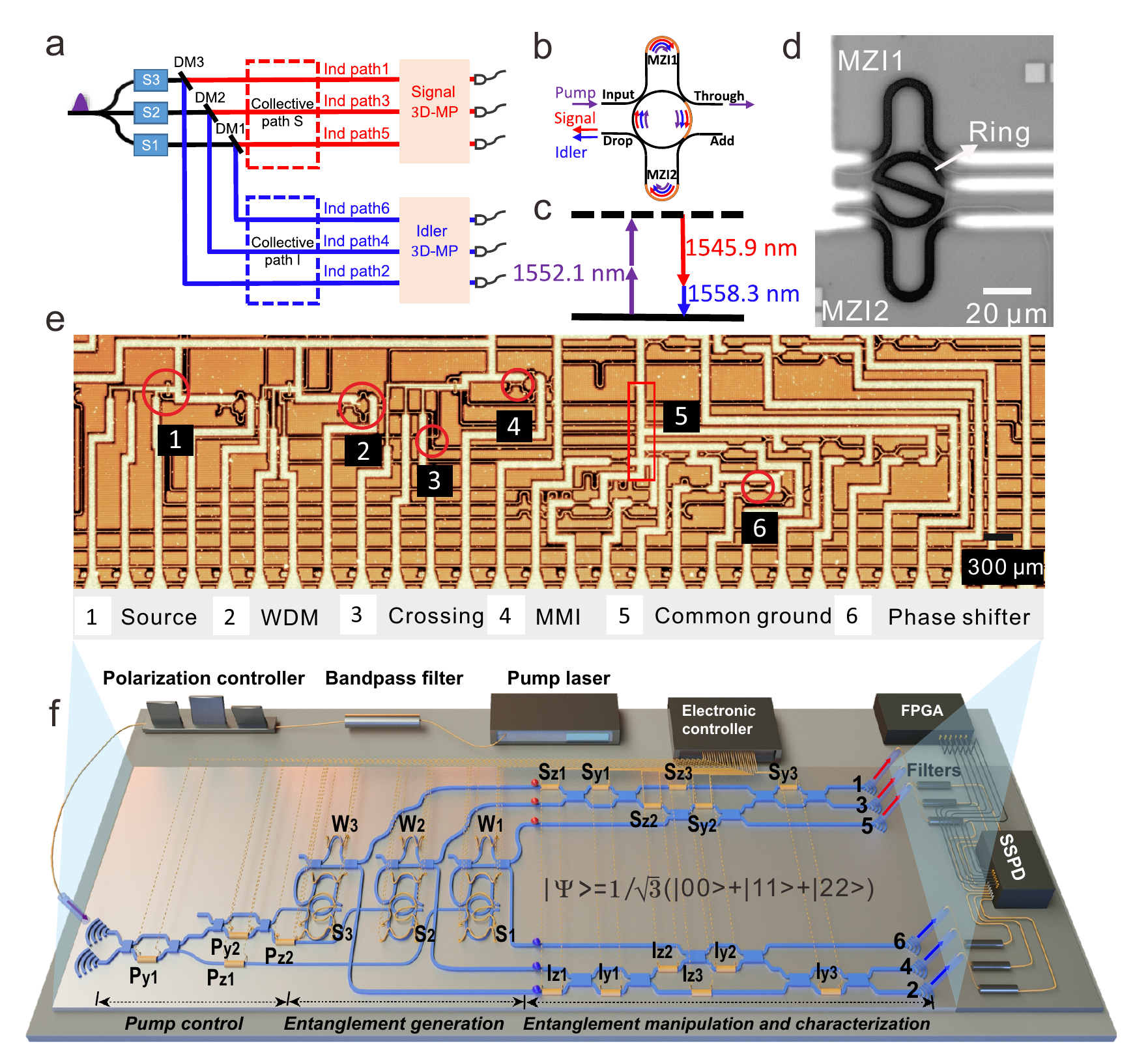}
    \caption{\label{Fig1} \textbf{Silicon quantum photonic chip and schematic of the experimental setup.} \textbf{a}, Conceptual scheme of the approach for generating and manipulating two entangled qutrits. Three non-degenerate photon-pair sources, S1, S2 and S3, are coherently pumped, and three dichroic mirrors, DM1, DM2 and DM3, separate signal and idler photons into two collective paths S and I, which consist of three individual (Ind) paths 1,3 and 5, and 2,4 and 6, respectively. Then two three-dimensional multiport interferometers (3D-MP) perform unitary transformations to the entangled qutrits. \textbf{b}, Diagram of a dual Mach-Zehnder interferometer micro-ring (DMZI-R) photon-pair source. This source has four ports: input, throughput, add and drop. Pump photons are sent into the input port and critically coupled into the ring resonator, where signal and idler photons are generated. By adjusting the phases of MZIs 1 and 2, we can reduce the photon leakage into the through port and increase the signal and idler photons coupling to the drop port. \textbf{c}, Two identical pump photons (1552.1 nm) generate one signal ( 1545.9 nm) and one idler (1558.3 nm) photons in non-degenerate spontaneous four wave mixing. \textbf{d}, Optical microscopy image of the DMZI-R photon-pair source. Thermo-optical phase shifters (PSs) are shown as thick black curves. \textbf{e}, Optical microscopy image of the whole entangled qutrit chip. Several important on-chip elements are labelled: 1. DMZI-R photon-pair source; 2. wavelength division multiplexer (WDM); 3. crossing; 4. multi-mode interferometer (MMI); 5. common ground of the electrical signals; 6. thermo-optic PS. \textbf{f}, Schematic of the complete experimental setup. A picosecond pump pulse is filtered, polarized and coupled into the chip via a bandpass filter, a polarization controller and a grating coupler, respectively. A photon pair is created in a superposition between three coherently pumped DMZI-R sources (S1-3). By adjusting four PSs for the pump (P$_{y1}$, P$_{y2}$, P$_{z1}$ and P$_{z2}$), we can generate a tunable qutrit entangled state. The signal (red) and idler (blue) photons are separated by W1-3 and routed to two 3D-MPs, which are composed of 12 MMIs and 12 PSs and enable us to implement arbitrary 3D local unitary transformation. Both signal and idler photons are then coupled out from the chip, filtered and detected by six grating couplers, filters and superconducting single photon detectors (SSPDs). Coincidence events are recorded by an FPGA-based timetag unit. All on-chip PSs are controlled with current sources.}
\end{figure}

 We have employed a scalable scheme for generating high-dimensional entangled states~\cite{schaeff2012}. As shown in the conceptual scheme (Fig.1 \textbf{a}), entangled qutrits are generated by three coherently pumped non-degenerate spontaneous four wave mixing (SFWM) photon-pair sources, in which two pump photons generate one signal photon and one idler photon with different wavelengths (Fig.1 \textbf{b} and \textbf{c}). The signal and idler photons are separated by dichroic mirrors (DMs) and then sent through reconfigurable linear optical circuits for implementing arbitrary 3-D unitary operations via three-dimensional multiports (3D-MPs). Finally, we verify and harness the qutrit entanglement by detecting single photons at the outputs.

To obtain an efficient photon-pair source, we use a dual Mach-Zehnder interferometer micro-ring (DMZI-R) photon-pair source~\cite{tison2017, vernon2017}. Such a DMZI-R photon-pair source is inspired by the design of a wavelength division multiplexer (WDM) in classical optical communication~\cite{barbarossa1995} and could circumvent the trade-off between the utilization efficiency of the pump photon and the extraction efficiency of the signal and idler photon pairs from a ring resonator~\cite{vernon2016}. The DMZI-R photon pair source was first demonstrated in ref.~\cite{tison2017}, where enhanced coincidence efficiency of a single source was shown. The working principle of the DMZI-R photon source is as follows: by wrapping two pulley waveguides around a ring resonator, one can construct a four-port device, as shown in Fig. 1\textbf{b}. We couple each waveguide at two points to the resonator, using four directional couplers. By adjusting the relative phases of two waveguides to the resonator, we can tune the coupling between waveguides to the resonator at the pump, signal and idler wavelengths independently. In the ideal case, we would like to have the pump circulate in the ring resonator to generate photon pairs, and therefore, the critical coupling condition for the pump is preferred. On the other hand, we want to extract the signal and idler photons from the resonator as soon as they are generated to minimize the propagation loss of photon pairs in the resonator. As a consequence, it is desirable to over-couple the waveguide to the resonator at the wavelengths of signal and idler photons. These two requirements can be fulfilled simultaneously by setting the free spectral range (FSR) of the MZIs to be twice that of the ring, such that every second resonance of the ring is effectively suppressed. By doing so, we can achieve the desired distinct coupling conditions for the pump, signal and idler photons and maximally utilize the pump to efficiently extract photon pairs. The ring has a radius of 15 $\mu$m and a coupling gap of 250 $\mu$m (200 $\mu$m) at the input (output) side. The length difference of the unbalanced MZI1 (MZI2) is 47.8 $\mu$m (48 $\mu$m). Optical microscopy images of the DMZI-R photon-pair source and the whole entangled qutrit chip are shown in Fig. 1\textbf{d} and \textbf{e}, respectively. Note that the sizes of the gaps of critical and over-critical couplings depend on the propagating loss of the photons in the ring. One should be able to obtain a higher count rate by optimizing the gap size~\cite{tison2017} (see Supplementary Information for a theoretical analysis and a detailed characterization of the DMZI-R photon source).

Following the DMZI-R photon pair source, we use an asymmetric MZI (AMZI) as an on-chip WDM to separate the non-degenerate signal and idler photons. As shown in Fig. 1\textbf{f}, we repeat this combination of DMZI-R source and WDM three times, and excite these sources coherently. When the generation rate is the same for all three sources and the relative phases of the pump are all zero, we generate a maximally entangled state of two qutrits: $\ket{\Psi}=\frac{1}{\sqrt{3}}(\ket{00}+\ket{11}+\ket{22})$. Note that $\ket{0}$, $\ket{1}$ and $\ket{2}$ are the individual path states of single photons.

Each qutrit can be locally manipulated by a 3D-MP~\cite{Reck:1994wd}, which is composed of thermo-optic phase shifters (PSs) and multi-mode interferometers (MMIs). In particular, one of the essential components, formed by a single PS and a tunable beam splitter, is realized with an MZI, consisting of two balanced MMIs and a PS.These components are used to realize R$_z({\varphi}_z)$ and R$_y({\theta}_y)$ rotations, and thus to obtain an arbitrary SU(2) operation in the two-dimensional subspace. Note that our experimental configuration is also closely related to a recent proposal on generating OAM entanglement by path identity~\cite{krenn2017entanglement}. The collective paths and individual paths in our work correspond to the path and OAM in ref.~\cite{kysela2019}. After manipulating and characterizing the entangled qutrits with two 3D-MPs, both the signal and idler photons are coupled out from the chip, filtered to suppress residual pumping with off-chip filters, and detected with superconducting nanowire single-photon detectors (SSPDs). The single-photon detection events are recorded by a field-programmable gate array (FPGA)-based time-tag unit. Then both single counts and coincidence counts CC$_{ij}$ between path i (i=1,3,5) and path j (j=2,4,6) are extracted from these timetag records (see Supplementary Information for further experimental details).

\textbf{From qubit entanglement to qutrit entanglement}

\begin{figure*}
\begin{center}
    \includegraphics[width=0.85\textwidth]{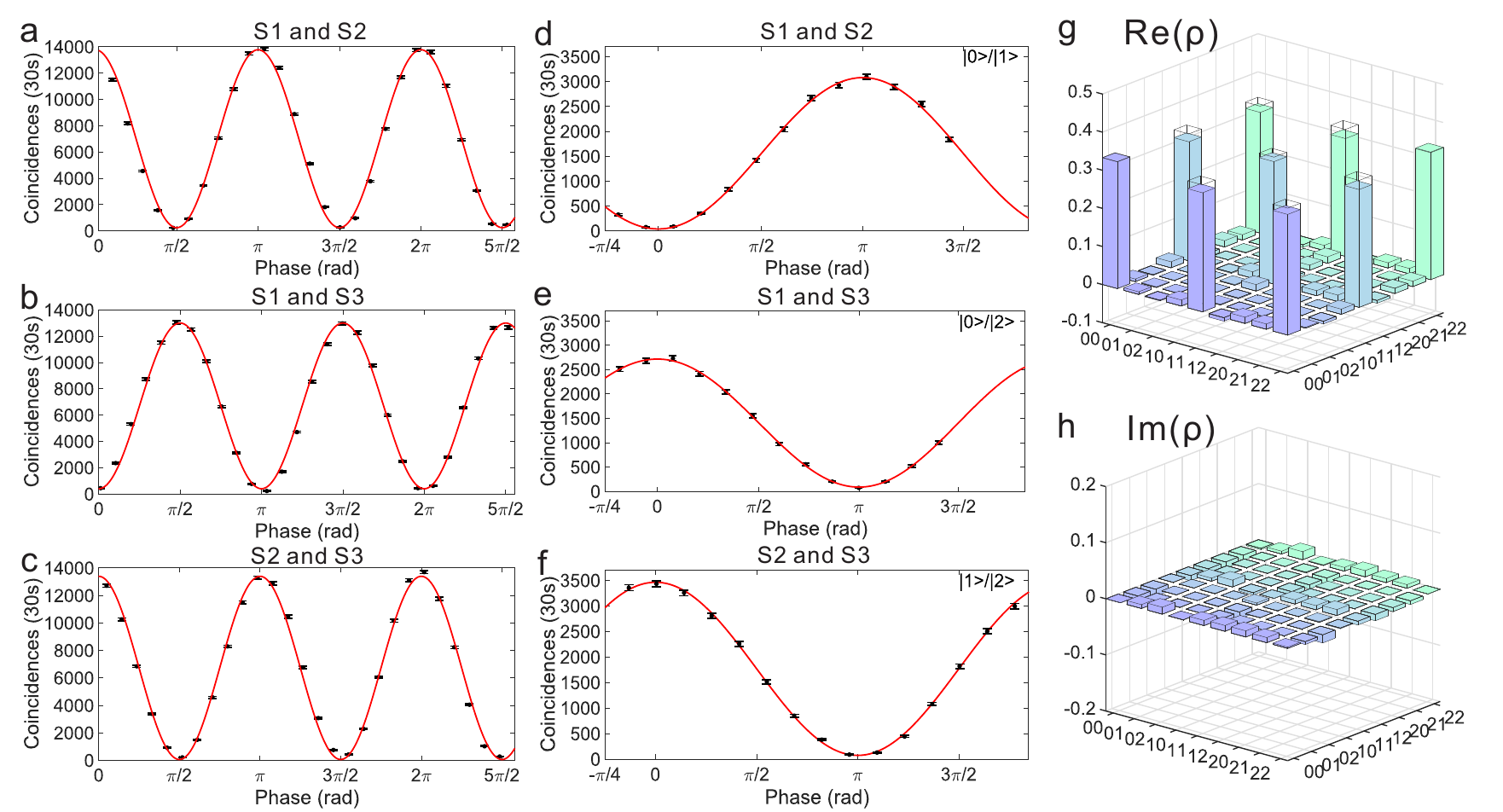}
    \caption{\label{Fig2} \textbf{Quantum interference and entanglement of two qutrits.} Interference fringes of the two-photon RHOM experiments are shown in \textbf{a}, \textbf{b} and \textbf{c} for S1 and S2, S1 and S3, and S2 and S3, respectively. The measured visibilities are greater than 96.49\% in all three cases, indicating high-quality indistinguishabilities between the three sources. Path-correlations of the three two-qubit subspaces for the entangled qutrit state $\frac{1}{\sqrt{3}}(\ket{00} +\ket{11} +\ket{22})$ are shown in \textbf{d}, \textbf{e} and \textbf{f} for S1 and S2, S1 and S3, and S2 and S3, respectively. The signal photon is projected on to $\frac{1}{\sqrt{2}}(\ket{j}+\ket{k})$, and the idler photon is measured in the base $\frac{1}{\sqrt{2}}(\ket{j} +e^{i\varphi }\ket{k})$ with $\varphi$ being the scanning phase and with (j,k)= (0,1),(0,2) and (1,2), respectively. The measured visibilities are greater than 94.72\% in all three cases, indicating high-quality path entanglement. The points are experimental data, and the curves are fits. The uncertainties denote the standard deviations from the Poisson distribution of the raw photon counts. The real and imaginary parts of the reconstructed density matrix of the two-photon entangled qutrit state are shown in \textbf{g} and \textbf{h}. Eighty-one measurement settings are decomposed into $\ket{p_s} \otimes \ket{p_i}$, where $\ket{p_k} =\ket{\psi_k}\langle\psi_k|$. $\ket{\psi_k}$ is chosen from the following set: {$\ket{0}$, $\ket{1}$, $\ket{2}$, $\frac{1}{\sqrt{2}}(\ket{0}+\ket{1})$, $\frac{1}{\sqrt{2}}(\ket{1}+\ket{2})$, $\frac{1}{\sqrt{2}}(\ket{0} +\ket{2})$, $\frac{1}{\sqrt{2}}(\ket{0}+i\ket{1})$, $\frac{1}{\sqrt{2}}(\ket{1} +i\ket{2} )$ and  $\frac{1}{\sqrt{2}}(\ket{0} +i\ket{2})$}. We obtain the quantum state fidelity of the measured quantum states to the ideal state as 95.50\%$\pm$0.17\%. The maximum matrix elements of the imaginary part is smaller than 0.015. The coloured bar graph is the experimental result and the wire grid indicates the expected values for the ideal case. The uncertainties in the fidelities extracted from these density matrices are calculated using a Monte Carlo routine, assuming Poissonian errors.}
\end{center}
\end{figure*}

To generate three-dimensional (3D) path-entangled photons, it is necessary to ensure that all three coherently pumped photon-pair sources are identical. This means that the emitted photon pairs from different sources should be the same in all DOFs, including polarization, spatial mode, count rate and frequency. For our chip-based system, we use single-mode waveguides, which automatically give us the same polarization states and spatial modes of photons from different sources. However, the count rate and frequency of the photons are not necessarily identical for different sources. To eliminate the count rate distinguishability, we can tune the pump power of the individual source. The last DOF is the frequency. In non-resonant broad band ($\mathrm{\sim}$nm to tens of nm) photon-pair sources, such as silicon nanowires, one can use off-chip narrow-band filtering to post-select identical spectra of different photons~\cite{qiang2018, wang2018}, which unavoidably reduces the count rate. In the resonant sources, such as our DMZI-R source, we can actively tune the resonance wavelengths of each individual source with PSs. In doing so, we obtain identical photons without sacrificing the photon count rate, which is particularly important for multi-photon high-dimensional experiments. However, aligning the frequency of the narrow-band photons generated from resonance-enhanced sources is challenging within the sub-mm foot-print of our chip. The reasons are follows: silicon has a relatively high thermo-optic coefficient. On the one hand, this feature of silicon is desirable for realizing reconfigurable photonic circuits by using thermo-optic PSs with low power consumption. On the other hand, it presents an experimental challenge to stabilize the frequency of the single photons generated from resonance-enhanced photon-pair sources under several distinct configurations of thermal PSs, due to thermal cross-talk ~\cite{carolan2019}. 

As the first step to generate 3D entanglement, we verify the identicality of two sources with time-reversed Hong-Ou-Mandel (RHOM) interference~\cite{chen2007}. Highly indistinguishable photons produce the high visibility of RHOM interference. We investigate the indistinguishability between all pairs of the three sources by interfering signal-idler photon pairs generated from S1, S2 and S3 on the top 3D-MP. For instance, we set the phases S$_{y1}$ and S$_{y2}$ to be $\pi $ and S$_{y3}$ to be $\pi /2$ and scan the phase of S$_{z3}$ to obtain the RHOM interference fringe between S2 and S3. The RHOM interference fringes between S1 and S2, S1 and S3, and S2 and S3 are shown in Fig.2\textbf{a}, \textbf{b} and \textbf{c}, respectively. The visibility of the fringe is defined as V=(CC$_{max}$-CC$_{min}$)/(CC$_{max}$+CC$_{min}$), where CC$_{max}$ and CC$_{min}$ are the maximum and minimum of the coincidence counts. The measured visibilities are greater than 96.49\% in all cases, indicating high-quality spectral overlaps. All of our data are raw, and no background counts are subtracted. To obtain high interference visibilities, we have spent significant amount of efforts to eliminate thermal noise (see Supplementary Information for further experimental details). We believe that by using better designs of the thermal PSs such as reported in ref.~\cite{harris2014,gao2019}, the noise can be greatly mitigated. In additional to reduce cross-talks, for reaching a visibility required for the practical applications, high-fidelity quantum control is a necessity. Remarkably, previous work has shown one can achieve excellent on/off ratio ($\sim$60.5 dB), equivalent to having a Pauli-Z error rate of $<10^{-6}$~\cite{wilkes2016} by using cascaded MZIs. Over 100 dB pass-band to stop-band contrast filters have also been realized by cascaded microrings~\cite{ong2013} and AMZIs~\cite{piekarek2017}, respectively. By combining these high-performance devices, we believe integrated quantum photonics is a promising route in the development of future quantum technologies and applications~\cite{rudolph2017}.

The next step is to verify the qubit entanglement of path states between three different pairs of sources. We measure the correlation of path entangled states in mutually unbiased bases (MUBs). We set the measurement base of the signal photon to be the coherent superpositions of the computational base, $\frac{1}{\sqrt{2}}(\ket{j} +\ket{k})$, where (j,k)= (0,1),(0,2),(1,2). Then, we scan the phase $\varphi$ in the quantum state of the idler photon, $\frac{1}{\sqrt{2}}(\ket{j} +e^{i\varphi}\ket{k})$ and measure the coincidence counts between the signal and idler photons. The coincidence fringes are shown in Fig. 2\textbf{d}, \textbf{e} and \textbf{f} for S1 and S2, S1 and S3, and S2 and S3, respectively. The values of various visibilities range from 94.72\%$\mathrm{\pm}$0.50\% to 97.50\%$\mathrm{\pm}$0.38\%, indicating high-quality qubit entanglement. The phase doubling signature of RHOM fringes compared to the correlation counterparts can be seen by comparing Fig. 2\textbf{a} \textbf{b} and \textbf{c}, and Fig. 2\textbf{d}, \textbf{e} and \textbf{f}. In the RHOM experiment, both signal and idler photons create a coherent superposition of two photons in two paths, the state that evolves under a phase shift in one of the modes then displays phase doubling~\cite{chen2007}. Note that the count rate of the path-correlation measurement is lower than that of the RHOM measurement, mainly because the design of the on-chip WDM is not optimal. Higher count rates of the correlation measurement can be achieved by optimizing the length difference between the two arms of the WDM.

Having established high-quality qubit entanglement, we proceed to characterize the qutrit entangled state with complete high-dimensional quantum-state tomography (QST). We use a set of all possible combinations of Gell-Mann matrices and apply the corresponding settings to both 3D-MPs~\cite{thew2002}. The QST method takes approximately 33 mins with a typical count rate $\mathrm{\sim}$100Hz per setting. Figure 2\textbf{g} and \textbf{h} display the real and imaginary parts of the reconstructed density matrix of the state, respectively, showing good agreement between the maximally entangled and measured quantum states with a fidelity of 95.50\%$\pm $0.17\%. The maximum matrix element of the imaginary part is smaller than 0.015. From the reconstructed density matrix, we obtain an \textit{I}-concurrence of 1.149$\mathrm{\pm}$0.002 ~\cite{fedorov2011}. The uncertainties in state fidelity extracted from these density matrices are calculated using a Monte Carlo routine assuming Poissonian statistics. In the context of quantum communication, a multi-dimension entanglement-based Ekert91 QKD protocol~\cite{ekert1991} was initially proposed and analyzed in refs.~\cite{cerf2002,bruss2002}, where high-dimension mutually unbiased bases¡¯ correlations between two Alice and Bob can be used to generate keys. The upper bound error rate (ER) that guarantees security against coherent attacks for device-dependent QKD in three dimensions is 15.95\%. For a maximally qutrit entangled state, the fidelity (F) of the state can be used to infer the ER if Alice and Bob use the same MUB~\cite{zhu2019}; that is F=(3-4$\ast$ER)/3. From the fidelity we obtain, the ER is only 3.375\%, which is considerably below the required bound, indicating the high quality of our qutrit state.

\textbf{Tests of quantum nonlocality and contextuality with entangled qutrits}

To benchmark the high-quality qutrit entanglement and high-precision quantum control, we demonstrate experimental tests of quantum nonlocality and quantum contextuality. Violations of Bell inequalities based on local realistic theories provide evidence of quantum nonlocality. It has been demonstrated that, high-dimensional correlations compatible with local realism satisfy a generalized Bell-type inequality, the Collins-Gisin-Linden-Massar-Popescu (CGLMP) inequality, with I$_d\leqslant2\ $for all d$\geqslant2$~\cite{collins2002}. Expression I$_3$ is given by joint probabilities as
\begin{equation} \label{GrindEQ__2_}
\begin{split}
I_3=[P(A_1=B_1)+P(B_1=A_2+1)+P(A_2=B_2)+P(A_1=B_2)] \\
-[P(A_1=B_1-1)+P(B_1=A_2)+P(A_2=B_2-1)+P(B_2=A_1-1)],
\end{split}
\end{equation}
where P$\left(A_{\mathrm{a}}\mathrm{=}B_{\mathrm{b}}+k\right)$ with (a,b=1,2) and (k=0,1) represent the joint probabilities for the outcomes of $A_{\mathrm{a}}$  that differ from $B_{\mathrm{b}}$ by k. The measurement bases used to maximise the violation of Eq. $\eqref{GrindEQ__2_}$ for the maximally entangled state $\ket{\Psi}=\sum_{j=0}^{2}\ket{j}_A\bigotimes \ket{j}_B$ are defined as
\begin{equation} \label{GrindEQ__3_}
\ket{K}_{A,a}=\frac{1}{\sqrt{3}}\sum_{j=0}^2\exp[i\frac{2\pi}{3}j(K+\alpha_a)]\ket{j}_A,
\end{equation}
\begin{equation} \label{GrindEQ__4_}
\ket{L}_{B,b}=\frac{1}{\sqrt{3}}\sum_{j=0}^2\exp[i\frac{2\pi}{3}j(-L+\beta_b)]\ket{j}_B,
\end{equation}
where i=$\sqrt{-1}$, ${\alpha }_1=0$, ${\alpha }_2=1/2$, $ {\beta }_1=1/4$ and ${\beta }_2=-1/4$, $K$ and $L\in$ \{0, 1, 2\} denote Alice's and Bob's measurement outcomes respectively, and $\ket{j}$ denotes the computational basis. These measurement bases can be implemented by configuring PSs in the 3D-MPs. For instance, we set S$_{z1}$ =$\ 0.333\pi $, S$_{y1}$ =$0.5\pi $ , S$_{z2}$ =$0.583\pi$, S$_{y2}$ =$0.392\pi$ ,S$_{z3}$ =$ 0.779\pi $ and S$_{y3}$ =$0.5\pi$ to realize setting A$_{1}$. In the context of quantum computation, entanglement is the essential resource. For one-way quantum computation, ref.\cite{reimer2019} reported the noise sensitivity of a two-photon, three-level and four-partite (two DOFs) cluster state with entanglement witness. CGLMP inequality is an entanglement criterion with higher correlation requirements comparing to entanglement witness. We witness the existence of entanglement between two qutrits in our experiment by using CGLMP inequality. The classical bound is violated by 51.46$\sigma$ ($I_3=2.7307\pm0.0142$), benchmarking the resilience to errors. The experimental results for the four base settings are shown in Table~\ref{tab:1}.

\begin{table}[!htbp]
\centering
\caption{Detailed experimental results leading to a violation of the CGLMP inequality}
\label{tab:1}
\begin{tabular}{| l | c | c | c | }
	\hline
	 Probability & Result & Expected value \\ \hline
	P($A_{1}$=$B_{1}$) & 0.7664$\pm$0.0059  & 0.8293\\ \hline
 P($A_{1}$=$B_{1}$-1) & 0.1480$\pm$0.0050  & 0.1111\\ \hline
  P($A_{1}$=$B_{2}$) & 0.8098$\pm$0.0054  & 0.8293\\ \hline
   P($A_{1}$=$B_{2}$+1) & 0.1030$\pm$0.0042  & 0.1111\\ \hline
     P($A_{2}$=$B_{1}$-1) & 0.8292$\pm$0.0052  & 0.8293\\ \hline
 P($A_{2}$=$B_{1}$) & 0.1014$\pm$0.0041  & 0.1111\\ \hline
  P($A_{2}$=$B_{2}$) & 0.8011$\pm$0.0055  & 0.8293\\ \hline
   P($A_{2}$=$B_{2}$-1) & 0.1233$\pm$0.0045  & 0.1111\\
	\hline
\end{tabular}
\end{table}
Contextuality is a fundamental concept in quantum mechanics ~\cite{kochen1967, klyachko2008, kirchmair2009, lapkiewicz2011} and an important resource for fault-tolerant universal quantum computation ~\cite{howard2014}. A single qutrit is the simplest quantum system showing the contradiction between non-contextual hidden-variable models and quantum mechanics ~\cite{lapkiewicz2011}. However, the testability of the Kochen-Specker (KS) theorem is debated due to the finite precision in a single qutrit in practical experiments ~\cite{meyer1999,kent1999}. An approach based on maximally entangled qutrit pair has been proposed ~\cite{cabello2011}, which was recently realized with bulk optics  ~\cite{hu2016}.

The experimental setting is as follows. A pair of maximally entangled qutrits is sent to two parties, Alice (A) and Bob (B). Alice performs projective measurements, either $D^A_1$ or $T^A_0$, $T^A_1$, and Bob simultaneously performs measurement $D^B_0$, where $D^A_1$ and $D^B_0$ are dichotomic projectors with two possible outcomes, 0 or 1, and $T^A_0$ and $T^A_1$ are trichotomic projectors with three possible outcomes, $a$, $b$ or $c$. These four projectors are defined as $D^B_0=\ket{i}\langle i|,\ D^A_1=\ket{f}\langle f|, T^A_0=\ket{a_0}\langle a_0|$ and $T^A_1=\ket{a_1}\langle a_1|$, where $\ket{i}=(\ket{0}+\ket{1}+\ket{2})/\sqrt{3}$, $\ket{f}=(\ket{0}-\ket{1}+\ket{2})/\sqrt{3}$, $\ket{a_0}=(\ket{1}-\ket{2})/\sqrt{2}$ and $\ket{a_1}=(\ket{0}-\ket{1})/\sqrt{2}$. The non-compatibility loophole contextuality inequality can be expressed as ~\cite{cabello2011}:
\begin{equation} \label{GrindEQ__5_}
P(D^A_1=1|D^B_0=1)-P(T^A_0=1|D^B_0=1)-P(T^A_1=1|D^B_0=1)\leq 0,
\end{equation}
 where $P(D^A_1=1|D^B_0=1)$ stands for the conditional probability of Alice obtaining result $1$ for $D^A_1$ when Bob also obtains result $1$ with $D^B_0$. $P(T^A_0=1|D^B_0=1)$ and $P(T^A_1=1|D^B_0=1)$ are defined analogously. For our experiment, we need to reconfigure two 3D-MPs according to these projectors and measure the coincidence counts to reconstruct the conditional probabilities in Eq. $\eqref{GrindEQ__5_}$. For example, $\ket{f}$ can be projected to port 3 by setting S$_{z1}$ =$\ 0$, S$_{y1}$ =$0.5\pi $ , S$_{z2}$ =$ 1.25\pi $, S$_{y2}$ =$0.608\pi $, S$_{z3}$ =$\ 0$ and S$_{y3}$ =$\pi$. We experimentally violate the non-contextuality inequality by 9.5 standard deviations ($0.085\pm 0.009$). The detailed experimental results are listed in Table~\ref{tab:2}. The no-signalling conditions, confirming the compatibility assumption between the measurements of Alice and Bob, are checked, as shown in Table~\ref{tab:3}. The results deviate slightly from 0 because of experimental imperfections.
\begin{table}[!htbp]
\centering
\caption{Detailed experimental results for the test of the noncontextuality inequalities}
\label{tab:2}
\begin{tabular}{| l | c |c| }
	\hline
	 Conditional probability & Result & Expected value\\ \hline
 $P(D_1^A=1|D_0^B=1)$ & 0.1245$\pm$0.0080& 0.111 \\  \hline
 $ P(T_0^A=1|D_0^B=1)$ & 0.0253$\pm$0.0035& 0 \\  \hline
  $ P(T_1^A=1|D_0^B=1)$ & 0.0136$\pm$0.0027& 0 \\
	\hline
\end{tabular}
\end{table}
\begin{table}[!htbp]
\centering
\caption{No-signalling results between Alice and Bob}
\label{tab:3}
\begin{tabular}{| l | c | }
	\hline
	 Probability & Result\\ \hline
$|P(D_0^B=1|D_1^A, D_0^B)-P(D_0^B=1|T_0^A, D_0^B)|$& 0.0236$\pm$ 0.0089 \\ \hline
$|P(D_0^B=1|D_1^A, D_0^B)-P(D_0^B=1|T_1^A, D_0^B)|$& 0.0094$\pm$ 0.0088 \\ \hline
$|P(D_0^B=1|T_0^A, D_0^B)-P(D_0^B=1|T_1^A, D_0^B)|$& 0.0330$\pm$ 0.0087 \\
	\hline
\end{tabular}
\end{table}


\textbf{Harnessing two-qutrit quantum correlations: quantum simulation of graphs and quantum metrology}

High-order quantum correlations are unique properties of high-dimensional entangled quantum systems and are a central resource for quantum information processing. To probe the quantum correlations in the entangled qutrit system, we measure the coincidence counts between signal and idler photons under different MUBs by tuning both the phases of signal/idler and pump photons. Here we use the quantum correlation between two entangled qutrits to demonstrate the quantum simulation of graphs and quantum metrology based on a quantum multi-path interferometer with third-order non-linearity.

\noindent \textbf{Quantum simulation of graphs:} Graphs are mathematical structures for describing relations between objects and have been widely used in various areas, including physics, biology and information science. A graph typically consists of a set of vertices and edges connecting the vertices. A subset of the edges containing every vertex of the n-vertex graph exactly once is defined as a perfect matching of the graph. To find the number of perfect matchings of a graph is a problem that lies in the \#P-complete complexity class ~\cite{valiant1979}. To provide an algorithm to solve such a hard problem is highly desirable. Recent studies have shown that a carefully designed quantum optical experiment can be associated with an undirected graph ~\cite{krenn2017}. In Particular, the number of coherently superimposed terms of the generated high-dimensional quantum state from a quantum optical experiment is exactly the number of perfect matchings in the corresponding graph. Each vertex stands for an optical path occupied by a single photon and every edge represents a photon pair source. This scheme can be viewed as a quantum simulation of graphs.

\begin{figure*}[htbp]
\begin{center}
    \includegraphics[width=0.68\textwidth]{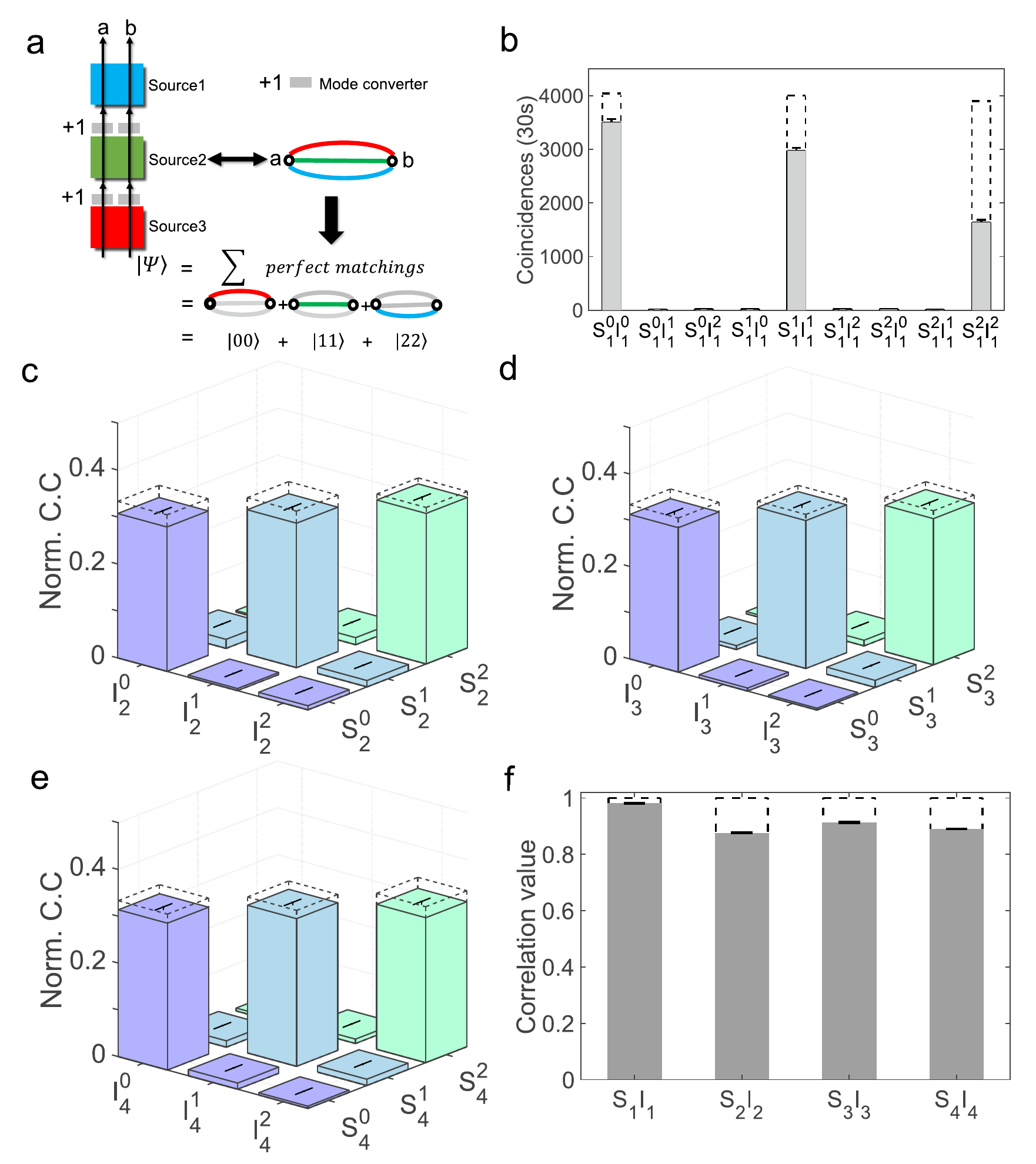}
    \caption{\label{Fig3} \textbf{Quantum simulation of graphs.} \textbf{a}, The optical setup corresponding to the experimental implementation on the chip and the corresponding graph. A two-fold coincidence in the experiment can be seen as a subset of edges containing each of the vertices only once, which is called a perfect matching in the graph. A coherent superposition of three perfect matchings leads therefore to the quantum state $\ket{\Psi}=\frac{1}{\sqrt{3}}(\ket{00} +\ket{11} +\ket{22}$.  \textbf{b}, The measured coincidence counts between the signal and idler photons on the computational basis, $S1I1$. Dashed empty bars are deduced results by balancing the loss for every port. \textbf{c-e}. The balanced normalized coincidence results for $S_{2}$$I_{2}$, $S_{3}$$I_{3}$ and $S_{4}$$I_{4}$. \textbf{f}, The experimental correlation coefficients are measured in all four MUBs ($S_{1}$$I_{1}$, $S_{2}$$I_{2}$, $S_{3}$$I_{3}$ and $S_{4}$$I_{4}$) for the entangled qutrits. Ideal results are shown with dashed empty bars. The uncertainties denote the standard deviations from the Poisson distribution for raw photon counts.}
\end{center}
\end{figure*}

As the first step towards the quantum simulation of graphs, we use entangled qutrits to experimentally demonstrate the connection between graph theory and quantum optical experiments. Figure 3\textbf{a} shows a conceptual scheme of our realization. Each pair of photons generated from sources propagates along their paths, denoted by black arrows, and acquires additional mode shifts due to the mode converters between the sources. As mentioned above, the path and OAM in Ref. ~\cite{krenn2017} are equivalent to the collective and the individual paths of our integrated quantum photonic circuit, as shown in Fig. 1\textbf{a}. Therefore, the mode converters can be implemented by routing the individual paths of the photon on our chip. By suitably setting up the 3D-MPs, we verify the resultant quantum state with coherent superimposed terms corresponding to the number of perfect matchings. We implement two experimental steps to realize this goal. First, we measure the coincidence counts between the signal and idler photons on the computational basis, S$_{1}$I$_{1}$. The experimental results are shown in Fig. 3\textbf{b}. It is clear that the major contributions are the 00, 11 and 22 terms. The second step is to verify the coherence between these three terms. For the entangled-qutrit pair system, each individual qutrit has four MUBs. Therefore, we have to measure the correlation coefficients in all four base combinations, i.e. S$_{1}$I$_{1}$, S$_{2}$I$_{2}$, S$_{3}$I$_{3}$ and S$_{4}$I$_{4}$, where S$_k=I_k^*$, with (k=1,2,3,4). These MUBs, up to normalization, are defined as
\begin{equation} \label{GrindEQ__6_}
\begin{split}
 S_1=(1, 0, 0), (0, 1, 0), (0, 0, 1),\\
  S_2=(1, 1, 1), (1, \omega, \omega^*), (1, \omega^*, \omega),\\
   S_3=(1,  \omega, 1), (1, \omega^*, \omega^*), (1, 1, \omega),\\
    S_4=(1, \omega^*, 1), (1, 1, \omega^*), (1, \omega, \omega),\\
    \end{split}
\end{equation}
where $\omega=e^{i\frac{2\pi }{3}}$ and * stands for the complex conjugation.

The normalized coincidence counts of S$_{2}$I$_{2}$, S$_{3}$I$_{3}$ and S$_{4}$I$_{4}$ are shown in Fig. 3\textbf{c}, \textbf{d} and \textbf{e}, respectively, by balancing the loss for every port. The experimental correlation coefficients derived from the coincidence counts in the four MUB combinations are shown in Fig. 3\textbf{f}. For an ideal maximally entangled-qutrit pair, the correlation coefficients should be unity. Due to the experimental imperfections, we obtain the correlation coefficients with the values of 98.17\%$\pm$0.11\%, 87.61\%$\pm$0.27\%, 91.32\%$\pm$0.24\% and 89.01$\pm$ 0.27\%, which shows correlations in all MUBs.

\begin{figure*}[htbp]
\begin{center}
    \includegraphics[width=0.85\textwidth]{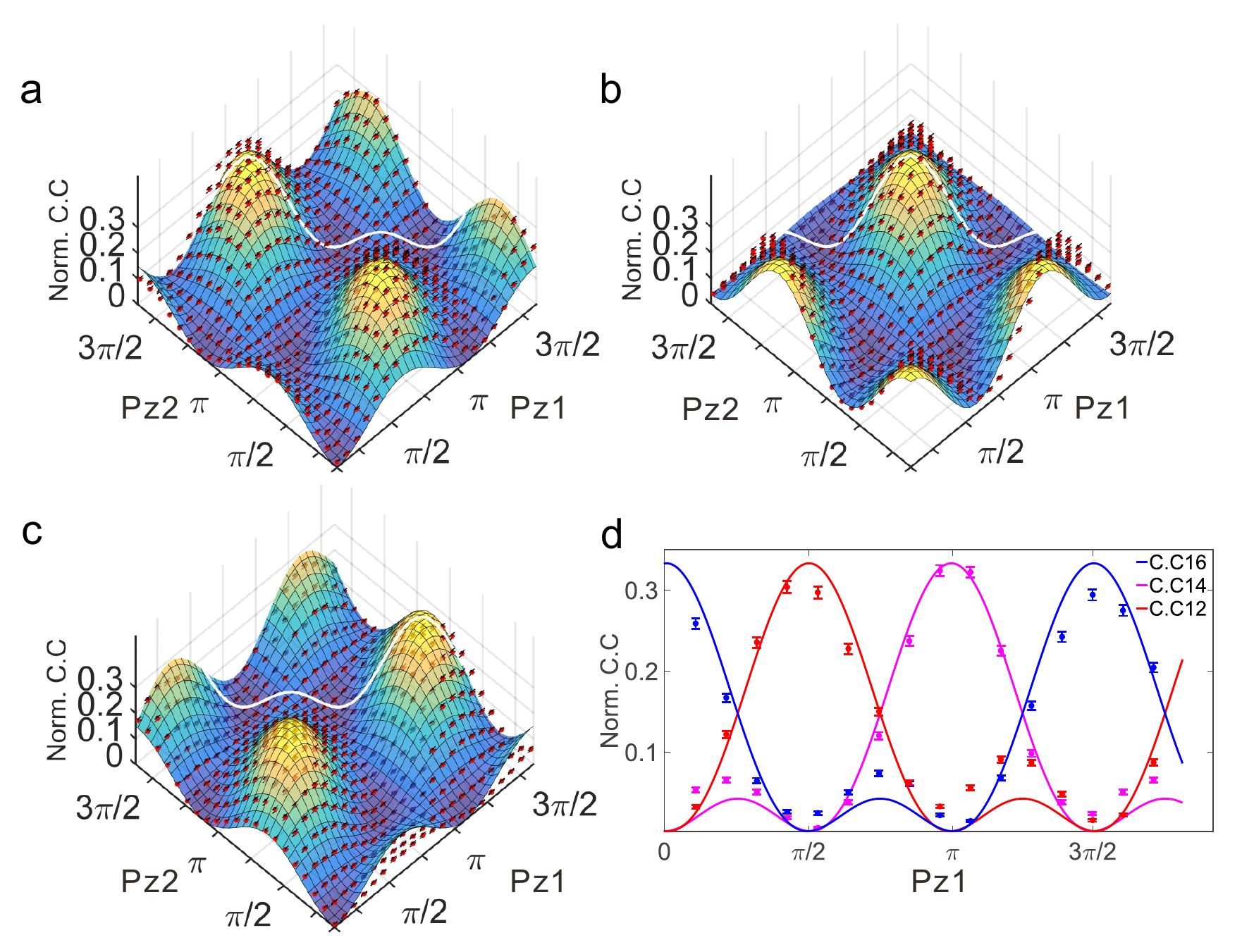}
    \caption{\label{Fig4} Quantum metrology with the entangled qutrits state $\ket{\Psi}=\frac{1}{\sqrt{3}}(e^{i2Pz1}\ket{00}+e^{i2Pz2}\ket{11}+\ket{22})$. \textbf{a-c} correspond to coincidence detection combinations of detectors (1,2) (1,4) and (1,6), respectively, with two relative pump phases Pz1 and Pz2 scanned. The normalized measured coincidences (Norm. C. C) (red dots) and simulated results without any free parameters (lines and surfaces) are shown. The data display excellent agreement with the theoretical predictions. \textbf{d}. Diagonal cuts through the three-dimensional plots by setting Pz1=-Pz2 in \textbf{a-c} reveal the typical structure of the three-path interferometer with a sensitivity of 1.476$\pm$0.048, exceeding both classical three-path linear interferometer and quantum second-order nonlinear limits. The red, pink and blue points represent the experimental data extracted from \textbf{a-c} and are identified as white lines respectively. The curves show the theoretical results.}
\end{center}
\end{figure*}

\noindent \textbf{Quantum Metrology with entangled qutrits:} Accurate phase measurements are at the heart of metrological science. One figure-of-merit for evaluating the accuracy of phase measurement is sensitivity, \textit{S}, which is defined as the derivative of the output photon number with respect to a phase change $S\propto 1/N$. In the experimental setting of classical interferometers, it is well known that increasing the number of paths of the interferometer can enhance the sensitivity ~\cite{sheem1981}. On the other hand, in the field of quantum metrology, one can further enhance the sensitivity by using entanglement ~\cite{giovannetti2011}. Here we combine both traits from classical and quantum systems and employ a three-dimension-entanglement third-order-nonlinearity interferometer to demonstrate the enhanced phase sensitivity compared to the classical three-path ~\cite{weihs1996} and quantum second-order-nonlinearity interferometers ~\cite{schaeff2015}. We send the entangled qutrit state $\ket{\Psi}=\frac{1}{\sqrt{3}}(e^{i2Pz1}\ket{00} +e^{i2Pz2}\ket{11} +\ket{22} $ into two separate 3D-MPs. We then scan the relative pump phases Pz1 and Pz2 and measure the coincidence between two qutrits (outputs 1,3,5 and 2,4,6). In total, there are nine different coincidence combinations. We quantify the qutrit-qutrit correlations as functions of phase settings of Pz1, Pz2. The normalized coincidences along with the theoretical results are shown in Fig. 4\textbf{a}-\textbf{c}. It is easily understood that the phase dependence is different between the second and third-order non-linear interactions ~\cite{reimer2016}. In the generation of entangled photon pairs, two pump photons are involved in the third-order processes providing a double phase compared to the second-order processes with the participation of only one pump photon. The measurements confirm this difference. Specifically, if the phases are chosen such that Pz1=-Pz2, the intensity varies from maximum to minimum as the pump phase is changed. This phase setting also gives the maximal sensitivity $S$. The raw data are extracted from the measured results and fitted with theoretical curves as shown in Fig. 4\textbf{d}. The averaged phase sensitivity $S=\frac{1}{CC_{abmax}}|\frac{dCC_{ab}}{d\varphi}|$ is $1.476\pm 0.048$ $rad^{-1}$, more than the theoretical ideal value of 0.5 $rad^{-1}$ for the two-path interferometer and 0.78 $rad^{-1}$ for the ideal three-path interferometer. The reason for the increased sensitivity is that the doubled phase sensitivity of the SFWM process and the side lobes appears between the two main peaks in three-path interference patterns, which enhance the steepness of the peaks of the correlations.
\section*{Discussion}
We have integrated three resonance-enhanced photon-pair sources embedded in interferometers, three WDMs and two 3D-MPs on a single monolithic silicon chip. We made all three sources identical without using frequency post-selection and observe high-visibility quantum interference, which allowed us to prepare, manipulate and analyse the high-quality path-entangled qutrit state. We violated the CGLMP inequality to confirm quantum nonlocality and the KS inequality to confirm contextuality with the entangled qutrits, verifying fundamental properties of quantum theory. Furthermore, we used two-qutrit quantum correlations to simulate graphs and identify the numbers of perfect matching for a small-scale graph. Finally, by using our chip for 3D entanglement, a third-order non-linearity interferometer, we improved the phase sensitivity by a factor of 2 compared to a classical three-path interferometer.

 Our demonstration of finding the number of the perfect matchings of the graph could be further extended to multi-photon and higher-dimension experiments, which might be suitable to demonstrate the quantum advantage in the near or mid term. To reach this regime, one needs to develop high-brightness multi-photon sources. The source presented in this work is a promising candidate for such a source.
 Although there exist a few technical challenges towards full integrated silicon quantum chips, such as cryogenic-compatible photon manipulation and high-efficiency photon detection, heterogeneous integrated chips are a promising approach for achieving this goal~\cite{he2019,ferrari2018,eltes2019}.
 Combined with an efficient on-chip SSPD~\cite{ferrari2018} and recently demonstrated cryogenic operation of Si-barium titanate~\cite{eltes2019}, our work could be viewed as a solid basis of future photonic quantum devices and systems for quantum information processing.

\section*{Data Availability}
The data that support the plots within this paper and other findings of this study are available from the corresponding author upon reasonable request.

\section*{Acknowledgements}
The authors thank M. Erhard, X. Gu, M. Krenn, A. Zeilinger and J. Wang for fruitful discussions. This research is supported by the National Key Research and Development Program of China (2017YFA0303704, 2019YFA0308704), National Natural Science Foundation of China (Grant No. 11674170, 11690032, 11321063, 11804153), NSFC-BRICS (No. 61961146001), NSF Jiangsu Province (No. BK20170010), the program for Innovative Talents and Entrepreneur in Jiangsu, and the Fundamental Research Funds for the Central Universities.

\section*{Competing interests}
The authors declare no competing interests.
\section*{Author contributions}
L.L., L.X., Z.C., and X.M. designed and performed the experiment. L.C., T.Y., T.T., Y.P. and X.C. provided experimental assistance and suggestions. W.M. provided theoretical assistance. L.L., L.X. and X.M. analysed the data. L.L. and X.M. wrote the manuscript with input from all authors. Y.L., S.Z. and X.M. supervised the project. L.L, L.X., Z.C. contributed equally to this work.

\newpage

\section{Supplementary Materials}
\textbf{Details of the silicon device and experiment}

\begin{figure*}[htbp]
\begin{center}
    \includegraphics[width=0.85\textwidth]{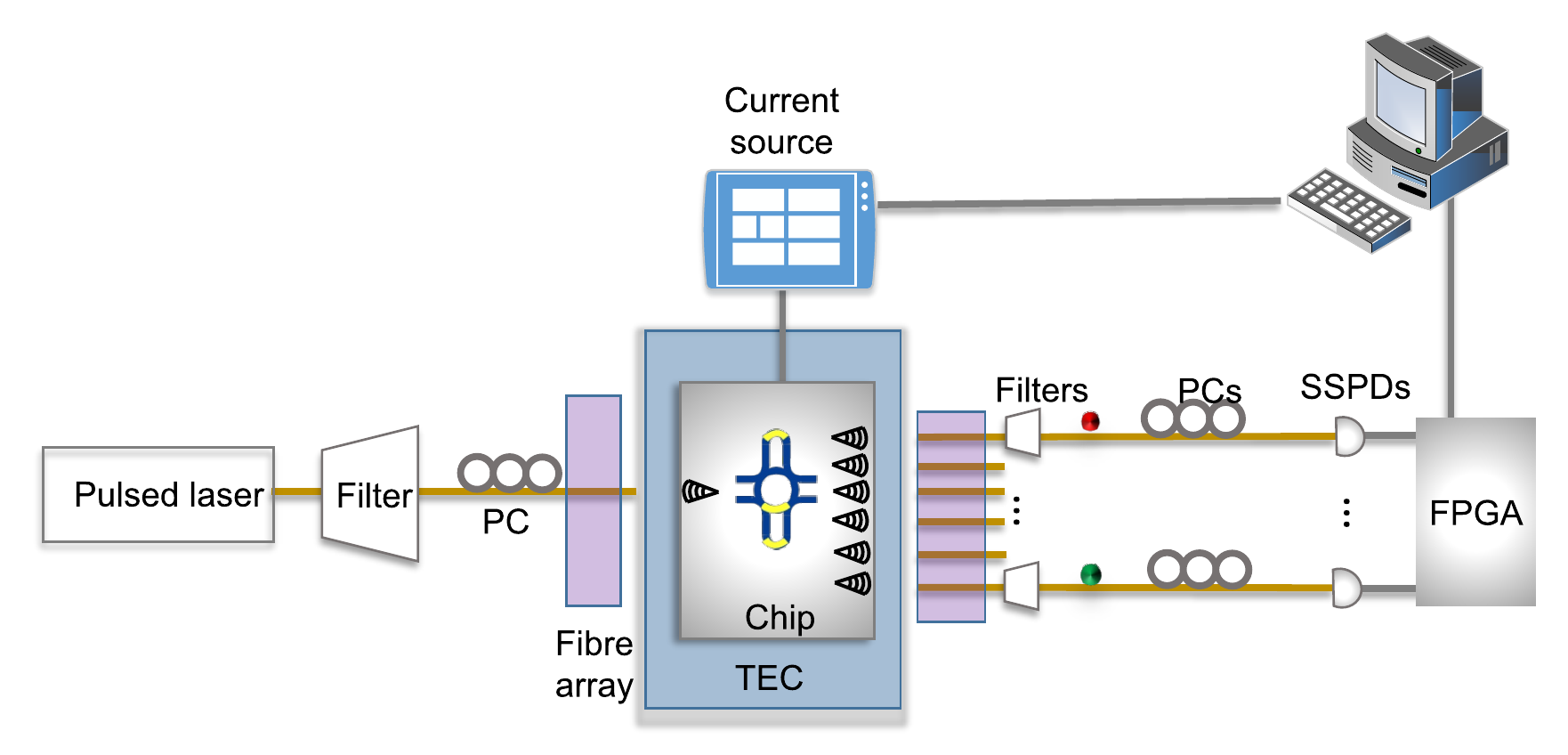}
    \caption*{Supplementary Figure 1. Schematic of the experimental setup. BS: beam splitter; PC: polarization controller; PID: proportional integrative derivative; SSPD: superconducting nanowire single photon detector; TEC: temperature controller; FPGA: field-programmable gate array.}
\end{center}
\end{figure*}
A schematic of the experimental setup is shown in Supplementary Figure 1. The components used in the experiment are all commonly used components for C-band wavelengths. Photon pairs are generated by coupling the pump light into the resonator using spontaneous four wave mixing (SFWM). Then the asymmetric Mach-Zehnder interferometer (AMZI) with a length difference of 47$\mu$m is used as an on-chip wavelength division multiplexer (WDM) to split the signal and idler photons into two multi-port interferometers. In our case, the free spectral range (FSR) is approximately 12 nm, which introduces a 3 dB filtering loss for both the signal and idler photons when they are divided into the two output ports of the AMZI in the path correlation measurements. In the time-reversed Hong-Ou-Mandel (RHOM) measurements, we do not use the WDM to separate the signal and idler photons. Therefore, the RHOM interference counts (interference on one side) are approximately four times larger than the correlation counts in the main text. After a local reconfigurable manipulation, the photons are coupled out of the chip for detection. The chip is mounted on a stage and wire-bonded on a printed circuit board (PCB) for  electrical contact.

The device is a silicon-on-insulator (SOI) chip fabricated using 248 nm deep-UV lithography at the advanced micro foundry (AMF) with 220 nm top thickness on 2 $\mu$m buried oxide. The waveguides are 500 nm wide and covered with a 2.8 $\mu$m silicon dioxide upper cladding. Resistive heaters are patterned as thermo-optic phase shifters (PSs) on a 120 nm thick TiN metal layer and placed 2 $\mu$m above the waveguide layer. To change the phase, we change the current propagating through the TiN metal layer to heat the waveguide and then change the refractive index. Multi-mode interferometers (MMIs) are used as beam-splitters with near 50:50 splitting ratio in the on-chip WDM and MZIs (see Fig. 1 of the main text).
\begin{figure*}[htbp]
\begin{center}
    \includegraphics[width=0.55\textwidth]{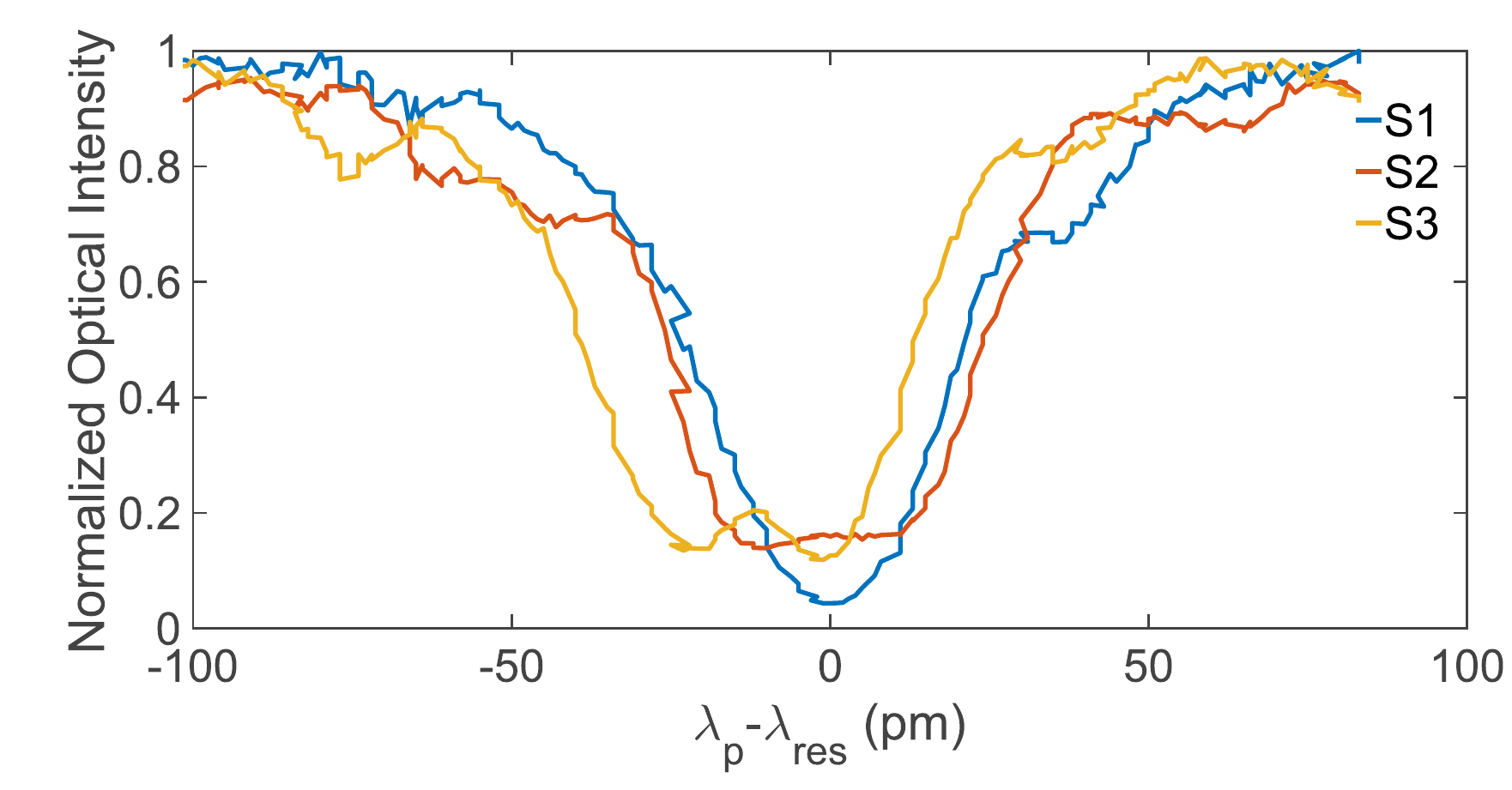}
    \caption*{Supplementary Figure 2. Spectral profiles of the micro-ring resonator sources. The full widths at half maximum (Q-factor) of $S1$, $S2$ and $S3$ are approximately 44 pm (3.5$\mathrm{\times }{10}^4$), 50 pm (3.1$\mathrm{\times }{10}^4$\textit) and 53 pm (2.9$\mathrm{\times }{10}^4$\textit), respectively. All the sources are observed to be largely spectrally indistinguishable.}
\end{center}
\end{figure*}
The photon pair generation source is designed with two pulley-type waveguides wrapped around the ring, essentially establishing two unbalanced MZIs. The ring has a radius of 15 $\mu$m and a coupling gap of 250 $\mu$m (200 $\mu$m) at the input (output) side. The length difference of the unbalanced MZI1 (MZI2) is 47.8  $\mu$m (48 $\mu$m) and its sinusoidal spectrum determines the wavelength of the constructive and destructive interference. Supplementary Figure 2 shows the pump resonance transmission spectra of source1 (S1), source2 (S2) and source3 (S3) with full widths at half maximum of 44 pm, 50 pm and 53 pm. The resonance splitting of S2 and S3 are due to surface-roughness-induced backscattering ~\cite{bogaerts2012}.

 To pump the sources, laser pulses are produced by a tunable picosecond fibre laser at 1552.02 nm (pulse duration of 7.8 ps, repetition rate of 60.2 MHz, and average power of -0.8 dBm). Before coupling the pump pulse into the chip, the unwanted amplified spontaneous emission (ASE) noise of the pump is filtered by a WDM (extinction ratio of $\sim$40 dB  and a bandwidth of 1.2 nm) and the polarization of the pump is optimized for maximum fibre-to-chip coupling by a polarization controller (PC). A 32-channel V groove fibre array with a spacing of 127 $\mu$m and polishing angle of $10^0$ is used to couple the light via transverse electric (TE) grating couplers into the chip. The FSR of the source is approximately 6.2 nm. Entangled photons emerging from the chip are filtered by off-chip single-channel WDMs to remove the residual pump and are finally detected by six superconducting nanowire single photon detectors (SSPD) with 80\% detection efficiency, 100 Hz dark count rates, a 50 ns dead time and a 100 ps timing jitter. The losses for the photons in the path entanglement measurement add up to 18.73 dB $\sim$ 19.13 dB, which can be decomposed as follows: the total fibre-chip-fibre loss is 15.56 dB (measured by launching non-resonance light of 1 mW and collecting the output power at the through port of the source), the insertion losses over the full band of the off-chip WDMs are 2-2.4 dB, and the detection losses are 0.97 dB. The detector electrical signals are collected by a field-programmable gate array (FPGA)-based timetag device. All the heaters are independently controlled by a programmable current source with a range of 0-20 mA and 16-bit resolution.

\textbf{Crosstalk}

\begin{figure}[htbp]
	\includegraphics[width=0.6\textwidth]{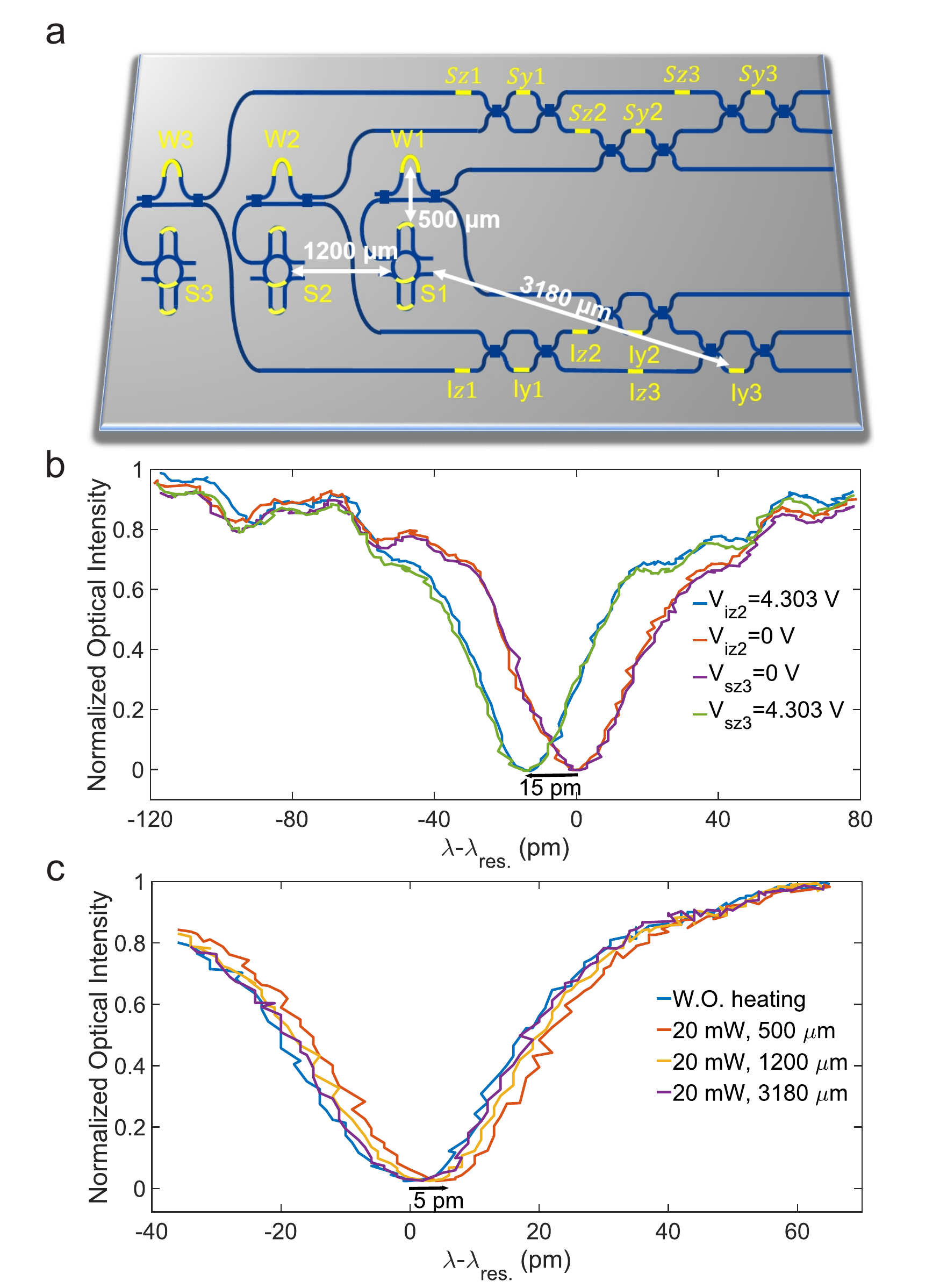}
	\caption*{Supplementary Figure 3. Characterization of the electric and thermal crosstalk of S1 on our chip. \textbf{a.} Schematic of the chip. \textbf{b.} The resonance of the source shifts 15 pm from its off position when 4.303 V is applied to the PS ($\varphi_{iz2}$  or $\varphi_{sz3}$). \textbf{c.} The thermal crosstalk is characterized by applying 20 mW of heating power to the PSs with different distances to the source. An obvious blueshift is observed for a small distance due to the heat dissipation.}
\end{figure}
Mitigating the electrical and thermal crosstalk within hundreds of micrometres is one of the main challenges in our experiment. The relative distances between each component are shown in Supplementary Figure 3a. To reduce the number of electrical pads, we designed all the heaters to have a common ground. However, the resistance between the common ground and true ground is not necessarily zero due to the electrical crosstalk. There are several ways to effectively avoid this effect such as a pre-processing method ~\cite{carolan2015}, a passive compensation method ~\cite{paesani2017}, a negative feedback scheme ~\cite{silverstone2015}, equipping each heater with two separate pads (one for the signal and one for the ground) and using current output equipment ~\cite{qiang2018}. We tried to use negative feedback to minimize the electrical crosstalk effect. However, as the number of heaters working simultaneously increases, it becomes increasingly difficult to optimize the feedback system to the precision. As shown in Supplementary Figure 3b, we measure the pump transmission spectrum of source 1 (S1) from the input port to the through port of the resonator with and without applying voltages to the other heaters. We choose two heaters ($I_{z2}$ and $S_{z3}$) with approximately equal resistances. When the voltage (4.303 V) is applied to $I_{z2}$ or $S_{z3}$, a common ground voltage ($V_G$) arises due to the resulting current passing through the contact resistance between the on-chip gold pad and wire-bonded contacts. The effective voltages (heating power) applied to the source are then reduced, resulting in a smaller effective refractive index ($n_{eff}$) change of the waveguide. The change in $n_{eff}$ as a function of temperature can be expressed as $\frac{dn}{dT}\Delta T$, where $\frac{dn}{dT}$ is the thermo-optic coefficient and $\Delta T$ is the change in temperature. The thermal-optic coefficient of silicon at 300 K near 1550 nm is $\mathrm{1.86}\mathrm{\times }{10}^{-4}K^{-1}$~\cite{harris2014}. Optical resonances in a resonator require that the optical length ($n_{eff}$L, where L is the round-trip length) is a multiple of the wavelength of the light. Hence, the reduction in $n_{eff}$ leads to the blueshift of the resonance wavelength. In Supplementary Figure 3b, the resonance wavelength blueshifts 15 pm when the voltage is applied to $I_{z2}$ or $S_{z3}$. The two approximately equal resistances with the same applied voltages provide similar electrical crosstalk, so the blue and green curves in Supplementary Figure 3b overlap well. We finally chose to use the current source, which is suitable for our experiment.

The thermal crosstalk originating from the heat transfer between the thermos-optical PSs also plays an important role. When we adjust the PSs of the WDMs and the multi-port interferometers, the thermal crosstalk poses a significant challenge for the spectral alignment of different sources due to the narrow resonance linewidths of our resonator-based sources. To overcome this challenge, we optimize the heat sink of the chip carrier and the design of the chip. We quantify the thermal crosstalk with and without (W.O.) heating the PSs with different distances to the source. The heating power is set to 20 mW. Supplementary Figure 3c shows that when the distance between the heater and the resonator decreases, there will be an obvious redshift due to the thermal crosstalk. As can be seen from the figure, the resonance of S1 shifts approximately 5 pm when the heater is positioned at a distance of 500 $\mu$m from S1, while for a distance of 1200 $\mu$m, there is little influence due to the thermal crosstalk. In our design, we distanced the PS that needs to be adjusted and the nearest source and achieved a desired isolation level. From the interference fringes shown in Fig. 2 in the main text, we note that the heat diffusion from the PSs has little effect on the sources within a power change of 0$\mathrm{\sim}$42 mW (2$\pi$ phase).

\textbf{Theoretical analysis and experimental characterization of the dual Mach-Zehnder interferometer micro-ring (DMZI-R) source on a chip}

\begin{figure}[htbp]
\begin{center}
	\includegraphics[width=0.8\textwidth]{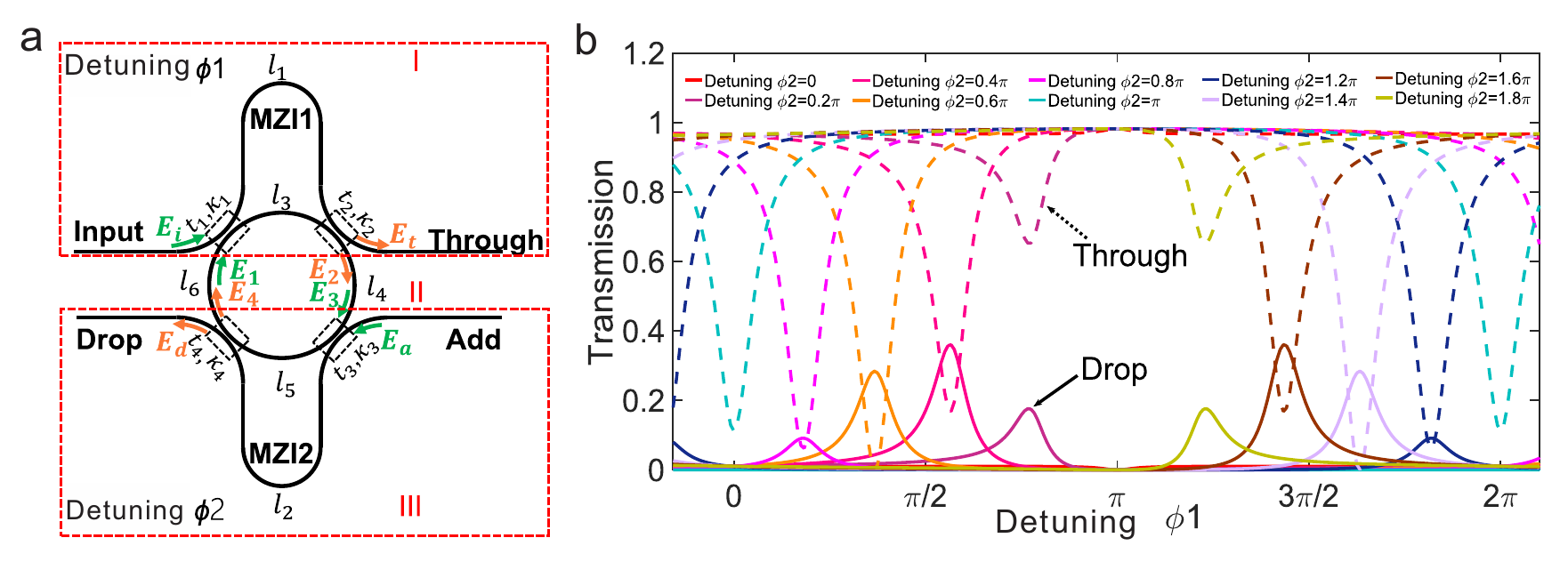}
	\caption*{Supplementary Figure 4. \textbf{a.} Schematic diagram of a DMZI-R. \textbf{b. }Theoretical calculations of the pump transmittance characteristics of the through port (dashed lines) and drop port (solid lines) with a attenuation coefficient $2\alpha=0.69$ $cm^{-1}$ for a ring with $r$=15 $\mu m$ versus of the detuning phase of the AMZI at the input side for different detuning phases of the AMZI at the output side.}
\end{center}
\end{figure}
We employ a quadruple-coupler ring resonator with the scheme presented in Supplementary Figure 4a, which is realized by wrapping two pulley waveguides around the input and output sides of the ring ~\cite{tison2017}. This configuration essentially establishes two unbalanced MZIs, MZI1 and MZI2, out of the waveguides and the ring. AMZI1 (AMZI2) consists of two directional couplers with coupling ratios of k1 and k2 (k3 and k4) and two arms with lengths of l1 and l3 (l2 and l5). For our chip, we set k1=k2 and k3=k4. The radius of the ring is r, while the lengths l4 and l6 are from coupler to coupler. We suppose that the directional couplers have zero length and a wavelength-independent coupling ratio within the wavelength range of interest. If the waveguide lengths are properly chosen, the FSR of the MZIs can be twice that of the ring.

The transmission spectrum of the resonator can be analysed by using a combination of transfer matrices ~\cite{barbarossa1995}. As shown in Supplementary Figure 4a, the resonators can be divided into two kinds of four-port blocks and two transmission waveguides, labelled as sections $I$, $II$ and $III$. For each section, two basic units, the directional coupler and the transmission waveguides, can be expressed in matrix form, relating the input ($E_{in1},E_{in2}$) and output ($E_{out1},E_{out2}$):
\begin{equation}\label{eq1}
\left( \begin{array}{c}
E_{out1} \\
E_{out2} \end{array}
\right)=H \cdot \left( \begin{array}{c}
E_{in1} \\
E_{in2} \end{array}
\right)=\left[ \begin{array}{c}
H_{11}\mathrm{\ \ \ }{\mathrm{H}}_{12} \\
H_{21}\mathrm{\ \ \ }{\mathrm{H}}_{22} \end{array}
\right]\cdot \left( \begin{array}{c}
E_{in1} \\
E_{in2} \end{array}
\right)
\end{equation}

For the directional coupler with a coupling ratio of $\kappa$, the transfer matrix is given as
\begin{equation}\label{eq2}
H_{\kappa}=\gamma\left[ \begin{array}{c}
t\mathrm{\ \ \ i}\kappa \\
i\kappa\mathrm{\ \ }t \end{array}
\right],
\end{equation}

where i=$\sqrt{-1}$, $\mathrm{t=}\sqrt{1-{\kappa }^2}$and $\gamma$ is the amplitude transmission coefficient of the coupler. For transmission waveguides with lengths of $l1$ and $l3$, we can define the transfer matrix as
\begin{equation}\label{eq3}
H_{l_1,l_3}=\left[ \begin{array}{c}
e^{i(\beta+i\alpha)l_1}\mathrm{\ \ }0 \\
0\mathrm{\ \ \ \ \ \ \ \ }e^{i(\beta+i\alpha)l_3} \end{array}
\right],
\end{equation}

\noindent where $\alpha$ is the waveguide transmission loss coefficient and $\beta$ is the propagation constant. Hence, we derive the transfer matrix $H_{C_1}$ for the cascaded components in section $I$ as
\begin{equation}\label{eq4}
H_{C_1}=H_{\kappa2}\cdot H_{l_1,l_3}\cdot H_{\kappa1},
\end{equation}
 which gives
\begin{equation}\label{eq5}
\left( \begin{array}{c}
E_t \\
E_2 \end{array}
\right)=H_{C_1}\cdot \left( \begin{array}{c}
E_i \\
E_1 \end{array}
\right).
\end{equation}

The same formulas can be used to derive the transfer matrix for section $III$, which relates the two input-output pairs ($E_a$,$E_3$) and ($E_d$, $E_4$)
\begin{equation}\label{eq6}
\left( \begin{array}{c}
E_d \\
E_4 \end{array}
\right)=H_{C_2}\cdot\left( \begin{array}{c}
E_a \\
E_3 \end{array}
\right),
\end{equation}
where
\begin{equation}\label{eq7}
H_{C_2}=H_{\kappa4}\cdot H_{l_2,l_5}\cdot H_{\kappa3}.
\end{equation}

The cascade of two-pair systems, section $I$ and section $III$, involves recirculating light; it is feasible to use a chain matrix G to combine the input-output pairs as
\begin{equation}\label{eq8}
\left( \begin{array}{c}
E_i \\
E_t \end{array}
\right)=G_{C_1}\cdot\left( \begin{array}{c}
E_2 \\
E_1 \end{array}
\right),
\end{equation}

and
\begin{equation}\label{eq9}
\left( \begin{array}{c}
E_4 \\
E_3 \end{array}
\right)=G_{C_2}^{-1}\cdot\left( \begin{array}{c}
E_a \\
E_d \end{array}
\right),
\end{equation}
where the elements of G can be derived from the transfer matrix H by the following conversion formulas ~\cite{moslehi1984}

\begin{equation}\label{eq10}
	\begin{split}
		G_{11} &=1/H_{21} \\
		G_{12} &=-H_{22}/H_{21}\\
        G_{21} &=H_{11}/H_{21} \\
        G_{22} &=(H_{12}H_{21}-H_{11}H_{22})/H_{21}.
	\end{split}
\end{equation}

Moreover, the relation between ($E_3$,$E_4$) and ($E_2$,$E_1$) in section $II$ is only related to the transmission lengths $l_4$ and $l_6$,
\begin{equation}\label{eq11}
\left( \begin{array}{c}
E_2 \\
E_1 \end{array}
\right)=T_{-l_4,l_6}\cdot\left( \begin{array}{c}
E_{4} \\
E_{3} \end{array}
\right)=\left[ \begin{array}{c}
0 \mathrm{\ \ \ } e^{-i(\beta+i\alpha)*l_4} \\
e^{i(\beta+i\alpha)*l_6} \mathrm{\ \ \ } 0 \end{array}
\right]\cdot \left( \begin{array}{c}
E_4 \\
E_3 \end{array}
\right)
\end{equation}

Substituting Eqs. $\eqref{eq9}$-$\eqref{eq11}$ in Eq. $\eqref{eq8}$, one can relate the mixed input-output pairs $(E_i,E_t)$ and $(E_d,E_a)$ as
\begin{equation}\label{eq12}
\left( \begin{array}{c}
E_i \\
E_t \end{array}
\right)=G_T\cdot\left( \begin{array}{c}
E_a \\
E_d \end{array}
\right),
\end{equation}

where
\begin{equation}
G_T=G_{C_1}\cdot T_{-l_4,l_6}\cdot G_{C_2}^{-1}.
\end{equation}

Finally, the transfer matrix $H_T$ from the input pair $(E_i,E_a)$ to the output pair $(E_d,E_t)$ can be determined from the elements of the chain matrix $G_T$ by using the following formulas:
\begin{equation}\label{eq13}
	\begin{split}
		H_{11} &=G_{21}/G_{11} \\
		H_{12} &=(G_{11}G_{22}-G_{12}G_{21})/G_{11}\\
        H_{21} &=1/G_{11} \\
        H_{22} &=-G_{12}/G_{11}.
	\end{split}
\end{equation}

By considering $E_a$=0, the transmittance characteristics can be obtained as
\begin{equation}\label{eq14}
E_d=\frac{\gamma^4(t_1\kappa_2\varphi_1+\kappa_1t_2\varphi_3)\varphi_4(\kappa_3t_4\varphi_2+t_3\kappa_4\varphi_5)}{1-\gamma^4(t_3t_4\varphi_5-\kappa_3\kappa_4\varphi_2)\varphi_6(t_1t_2\varphi_3-\kappa_1\kappa_2\varphi_1)\varphi_4},
\end{equation}

and
\begin{equation}\label{eq15}
E_t=\frac{\gamma^6(t_3t_4\varphi_1\varphi_3\varphi_4\varphi_5\varphi_6-\kappa_3\kappa_4\varphi_1\varphi_2\varphi_3\varphi_4\varphi_6)-\gamma^2(t_1t_2\varphi_1-\kappa_1\kappa_2\varphi_3)}
{\gamma^4(t_3t_4\varphi_5-\kappa_3\kappa_4\varphi_2)\varphi_6(t_1t_2\varphi_3-\kappa_1\kappa_2\varphi_1)\varphi_4-1},
\end{equation}
 where${\ \varphi }_i=e^{-\alpha l_i}e^{i{\theta }_i}$, with ${\theta }_i=\beta l_i$ (i=1,2,3,4,5,6). It can be seen from Eqs. $\eqref{eq14}$ and $\eqref{eq15}$ that the phase term $\varphi_i\ $has a critical influence on the resonator transmission characteristics. Without loss of generality, we assume that the four phase terms for the ring are equal (${\ \varphi }_3={\ \varphi }_4={\ \varphi }_5={\ \varphi }_6={\ \varphi }_r=\exp[-\alpha 2\pi r/4]\cdot\exp[i\beta 2\pi r/4]$). There are four distinct closed loops within the resonator, and the imposed resonance conditions for the round-trip phase delays are

\begin{equation}\label{eq16}
	\begin{split}
		{\theta }_1+{\theta }_2+2{\theta }_r &=N_1\times \pi\\
		{\theta }_1+{3\theta }_r &=N_2\times \pi \\
        {\theta }_2+3{\theta }_r &=N_3\times \pi  \\
        4\theta_r &=N_4\times\pi.
	\end{split}
\end{equation}

Theoretical calculations of the pump transmittance characteristic of the DMZI-R with $2\alpha=0.69$ $cm^{-1}$ and a radius of 15 $\mu$m are shown in Supplementary Figure 4b versus the detuning phase ${\phi }_1$ of AMZI1 for different detuning phases ${\phi }_2\ $ of AMZI2, where ${\phi }_1=\theta_1-\theta_r$ and ${\phi }_2=\theta_2-\theta_r$. The length difference of the AMZI1 (AMZI2) is 47.8 $\mu$m (48 $\mu$m), corresponding to the nominally designed parameters of our chip. The input (output) side directional coupler amplitude coupling ratio is set to $0.25$ $(0.2)$ in the absence of coupling loss ($\gamma $=1). It can be seen from this figure that the phase matching conditions rely on the interference pattern of the input and output MZIs. At the input side ${\phi }_1=2N\pi $ ensures that the pump laser couples into the resonator, while at the output side ${\phi }_2=(2N+1)\pi$ minimizes the pump leakage from the drop port and hence maintains the pump within the ring and isolates the ring from photons exiting via the drop port. The phase matching conditions for the generated photons (signal and idler) in the ring are exactly opposite.

The tunable relative phase between the different paths can be realized by the PSs in our device. By appropriately choosing the coupling ratio of the directional coupler and applying heaters to the device, one can effectively achieve critical coupling for the pump photon and over-coupling for both the signal and idler photons.
\begin{figure}[htbp]
\begin{center}
	\includegraphics[width=0.8\textwidth]{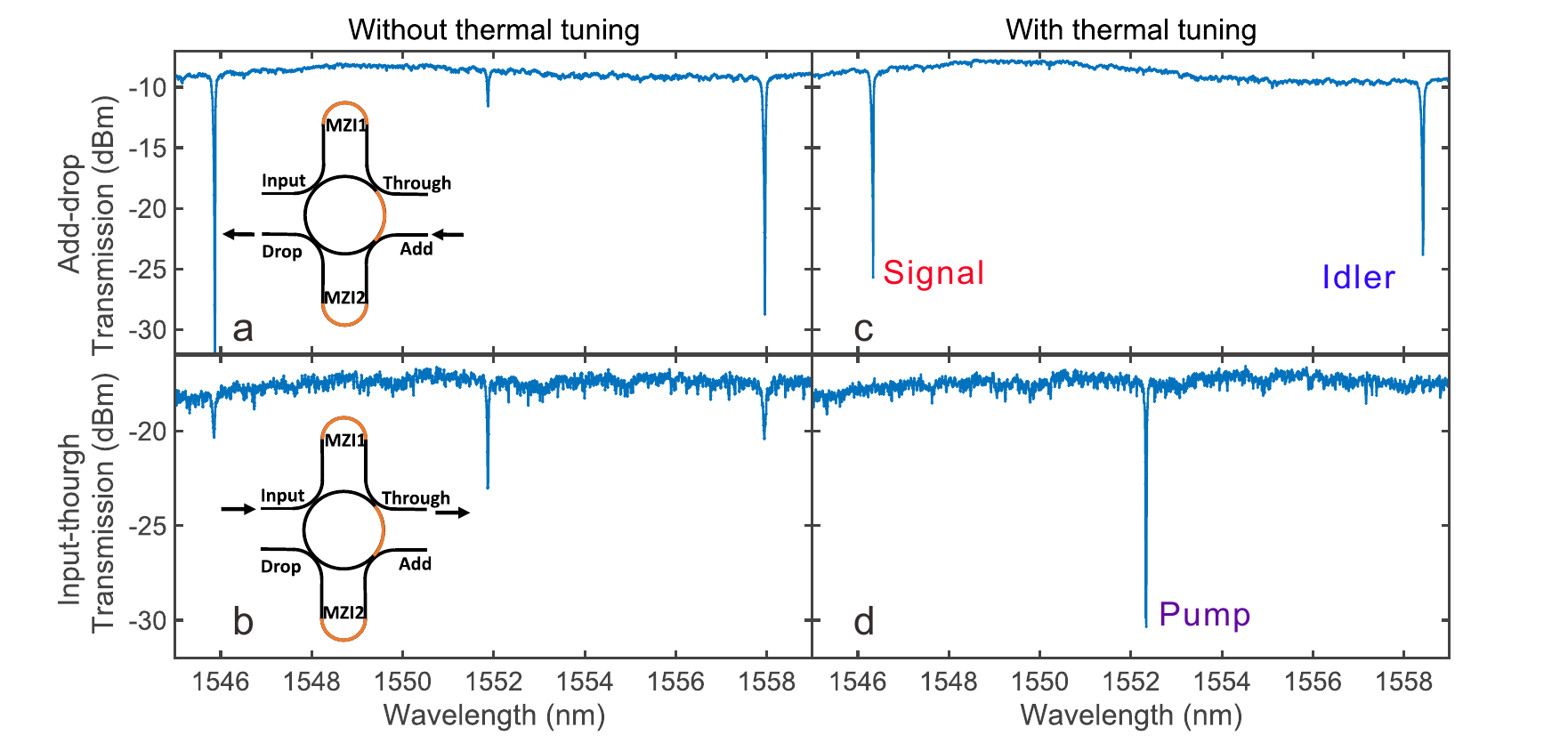}
	\caption*{Supplementary Figure 5. Spectrum characterization of the DMZI-R. Transmission spectra from the add port to the drop port and input port to the through port: \textbf{a.} add port to the drop port without thermal tuning; \textbf{b.} input port to the through port without thermal tuning; \textbf{c.} add port to the drop port with thermal tuning and \textbf{d. }input port to the through port with thermal tuning. The insets show a schematic diagram of the measurement.}
\end{center}
\end{figure}

Supplementary Figure 5 shows the transmission spectra of S1 from the add port to the drop port and from the input port to the through port before (a and b) and after (c and d) optimizing the heaters. From a comparison of Supplementary Figure 5a and 5c, we have enhanced the transmission from the add port to the drop port by tuning the phase of MZI2. Due to the complementary nature of the two-mode MZI, we can minimize the pump (indicated by "pump" in the figure) leakage from the ring into the drop port and hence maintain the pump within the ring for efficient SFWM. Moreover, at the wavelengths of the signal and idler photons (indicated with red and blue letters), the transmission increases; hence, the extraction efficiency of the correlated photon pair is also enhanced. On the other hand, from a comparison of Supplementary Figure 5b and 5d, by tuning the phase of MZI1, we reduce the transmission of the pump from the input port into the through port and hence maintain the pump within the ring for efficient SFWM. This design can greatly enhance the generation and extraction efficiency of the photon pairs simultaneously and reduce the amount of filtering needed for the pump. In addition, the design can also reduce the spurious SFWM pairs generated in the input waveguide if the coupler to the source is long and further improve the
signal-to-noise ratio.

\begin{figure}[htbp]
\begin{center}
	\includegraphics[width=0.65\textwidth]{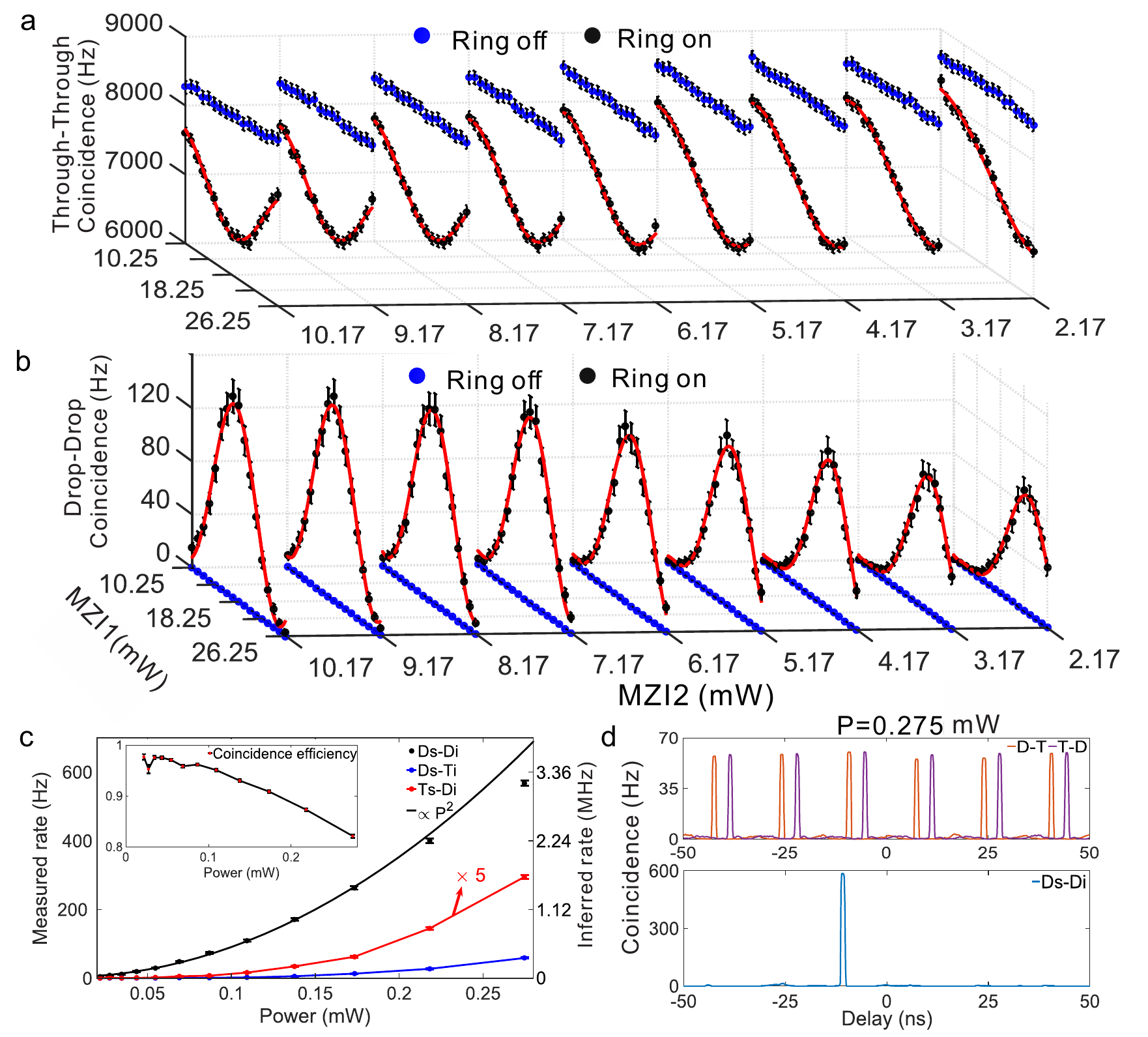}
	\caption*{Supplementary Figure 6. Characterization of S1. \textbf{a.} The through-through ports coincidence results of the on-resonance (black dots)
   and off-resonance (blue dots) cases upon performing a 2-D sweep of the current across the heaters above MZI1 and MZI2 with an average pump power of 0.13 mW before coupling into the chip. It is clear that in the on-resonant case, the coincidence counts depend strongly on the settings of the PSs of MZI1 and MZI2. However, the minimum of the interference fringe is high. Moreover, if we tune the ring to be off resonant with the pump, phase-independent coincidence counts are obtained, as shown by the blue data points. The high rate in the off-resonance case is a result of broadband SFWM in the input bus waveguide. \textbf{b.} 2-D sweep coincidence results of the drop-drop ports. In the off-resonance (blue dots) case, eliminating the possibility of photon pair generation within the resonator, the coincidence counts are approximately equal to zero. If the pump laser is tuned to be resonant (black dots), the settings of MZI1 and MZI2 have a significant effect on the drop port coincidences, depending on how efficiently the pump is coupled into the resonator.  \textbf{c.} Coincidence rate of different ports as a function of the average pump power. The dashed line is a quadratic fit. The inset shows that the coincidence efficiency changes with the average pump power. The points are experimental data, and the curves are fits. All the data are raw counts and no background counts are subtracted. The error bars are calculated by a Poissonian distribution. Since the Ds-Ti and Ts-Di lines are approximately overlapped, we multiply Ts-Di by five times to guide the eye. \textbf{d.} A typical histogram of the coincidence measurements at an input average pump power of 0.275 mW at different ports. This result shows the excellent signal-to-noise ratio of the DMZI-R source.}
\end{center}
\end{figure}

According to the optimum heater configuration that exhibits the desired spectral dependence shown in Supplementary Figure 5, we characterize the coincidence counts of S1 in detail. Supplementary Figure 6a shows the through-through (T-T) coincidence results upon performing a 2-D sweep of the current across the heaters above MZI1 and MZI2 for the on-resonance (black dots) and off-resonance (blue dots) cases, respectively. The average pump power is set to be 0.13 mW before coupling into the chip. We find that in the off-resonance case, the coincidence counts of T-T do not change with the modulation currents. The high rate in the off-resonance case is a result of broadband SFWM in the input bus waveguide, which will be discussed in the following paragraphs. On resonance, the coincidence counts strongly depend on the settings of MZI1 and MZI2 mainly because some of the pump light is coupled into the ring, resulting in less photon pair generation in the straight waveguide. In addition, the T-T coincidence counts cannot be reduced to zero because the photon pairs generated in the input waveguide before the source will be completely outcoupled from the through port when MZI1 is tuned to only support the pump photon at the input side. The drop-drop (D-D) coincidence counts are shown in Supplementary Figure 6b. Off-resonance, D-D dose not show a coincidence count since the pump is not coupled into the ring. For the on-resonance case, as the scan current gradually approaches the optimal configurations, the coincidence counts constantly increase, which complements the on-resonance case of T-T compared to Supplementary Figure 6a. Supplementary Figure 6c illustrates the coincidence counts of Ds-Di, Ds-Ti and Ts-Di as a function of the average pump power. The D-D coincidence rate increases with power, because the efficiency of SFWM scales quadratically with the pump power. The measured rate is 570 Hz at a pump power of 0.275 mW. Taking into account the losses of each photon (fibre-chip-fibre losses of 15.56 dB, insertion losses of the off-chip WDMs of 2.2 dB, and detection losses of 0.97 dB), the inferred generation rate is approximately 3.2 MHz in the resonator. The inset of Supplementary Figure 6c shows that the coincidence efficiency changes with the average pump power. To remove the effect of the photon generation from the input waveguide, we only use the drop port coincidences Ds-Di and split coincidences Ds-Ti and Ts-Di to
deduce the coincidence efficiency ~\cite{tison2017}:
\begin{equation}\label{eq17}
Coincidence \quad efficiency=\frac{CC^2_{Ds-Di}}{{({CC}_{Ds-Di}+\frac{{CC}_{Ds-Ti}+{CC}_{Ts-Di}}{2})}^2}.
\end{equation}
\noindent
As shown in the inset, in the case of low pump power, the coincidence efficiency is 0.97; however, as the power increases, i.e., as the coincidence rate of Ds-Ti and Ts-Di increases, the efficiency gradually decreases, which results from the accidental rate as shown in Supplementary Figure 6d. All the peaks in the histogram of the Ds-Ti and Ts-Di coincidences are similar, while in the case of D-D, the accidental peaks are almost not visible compared to the main peak. The peaks with a time interval of 16 $ns$ correspond to the simultaneous detection of photons from different pump pulses.

\begin{figure}
\begin{center}
	\includegraphics[width=0.8\textwidth]{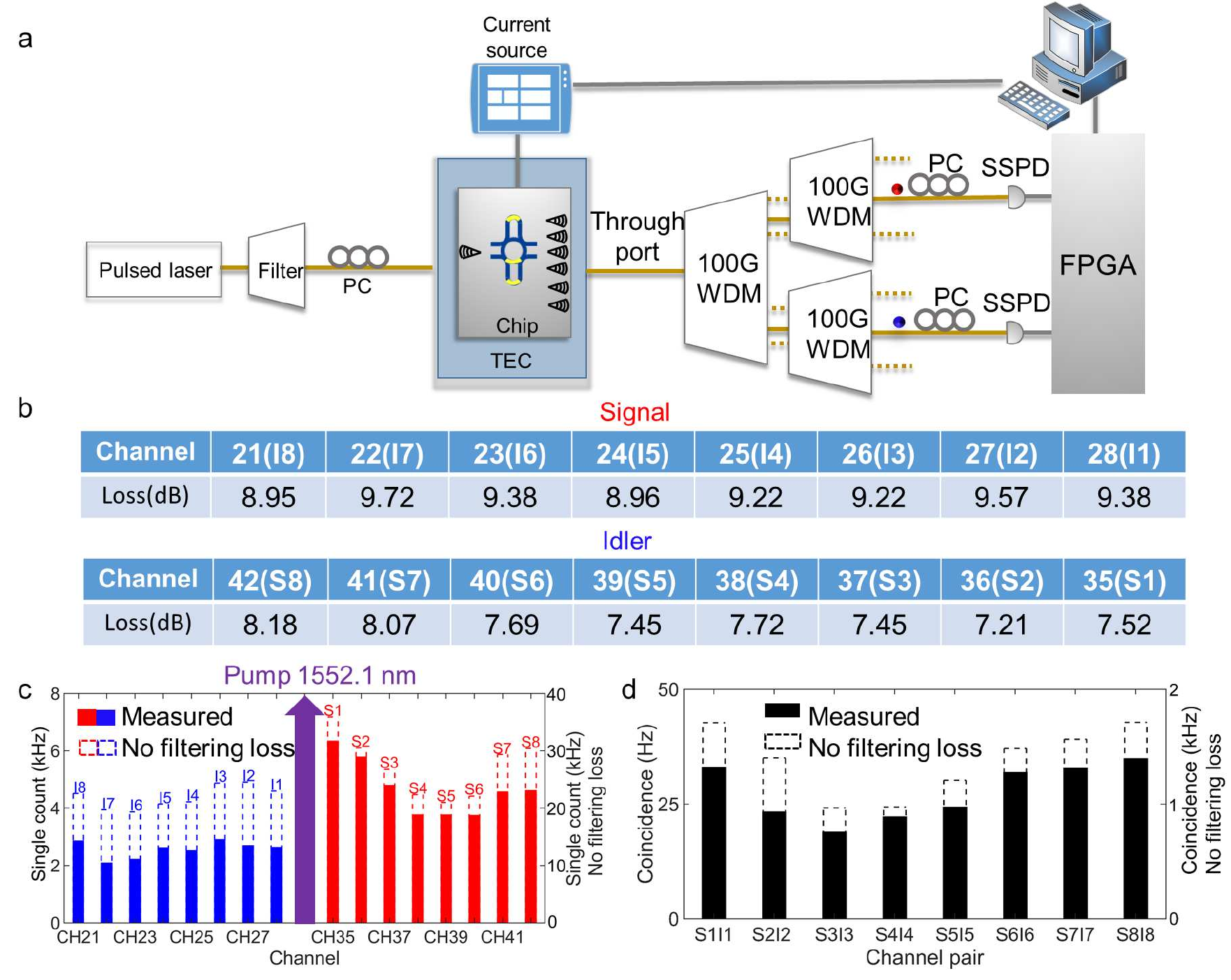}
	\caption*{Supplementary Figure 7. Characterization of the broadband SFWM in the input bus waveguide at the through port of the source. \textbf{a.} Schematic of the experimental setup. \textbf{b.} Measured losses of the multi-channel 100 GHz WDM. \textbf{c.} Single-count rate spectra of the signal and idler photons filtered by the WDM for the measured case and the case without filtering loss. \textbf{d.} The measured and inferred coincidence rates of the channel pair. }
\end{center}
\end{figure}

To illustrate the broadband SFWM in the bus waveguide, we measure the continuous emission spectra for the signal and idler photons with the currents set to the on-resonance conditions. In the experimental setup shown in Supplementary Figure 7a, a pulsed laser with an average power of 0.13 mW is first filtered by 200 GHz cascade dense WDM (DWDM) filters to remove the background noise and then is coupled into the chip. Photon pairs are generated from the chip through SFWM, and the photons coupling out from the chip are filtered by a multi-channel 100 GHz DWDM. Supplementary Figure 7c shows the single-count rates for different channels, with a maximum wavelength interval from the pump of approximately 8.5 nm. The rates without filtering loss are inferred based on the independently measured insertion losses for each channel shown in Supplementary Figure 7b. The differences in the single-count rates mainly arise from the Raman scattering of the signal and idler channels. The coincidence rates of the channel pairs are recorded and shown in Supplementary Figure 7d. It is clear that the spectrum of the SFWM in the input waveguide is substantially wider than that of the resonance-matched DMZI-R.

\textbf{Optimization of two-qubit entanglement}

\begin{figure}
\begin{center}
	\includegraphics[width=0.8\textwidth]{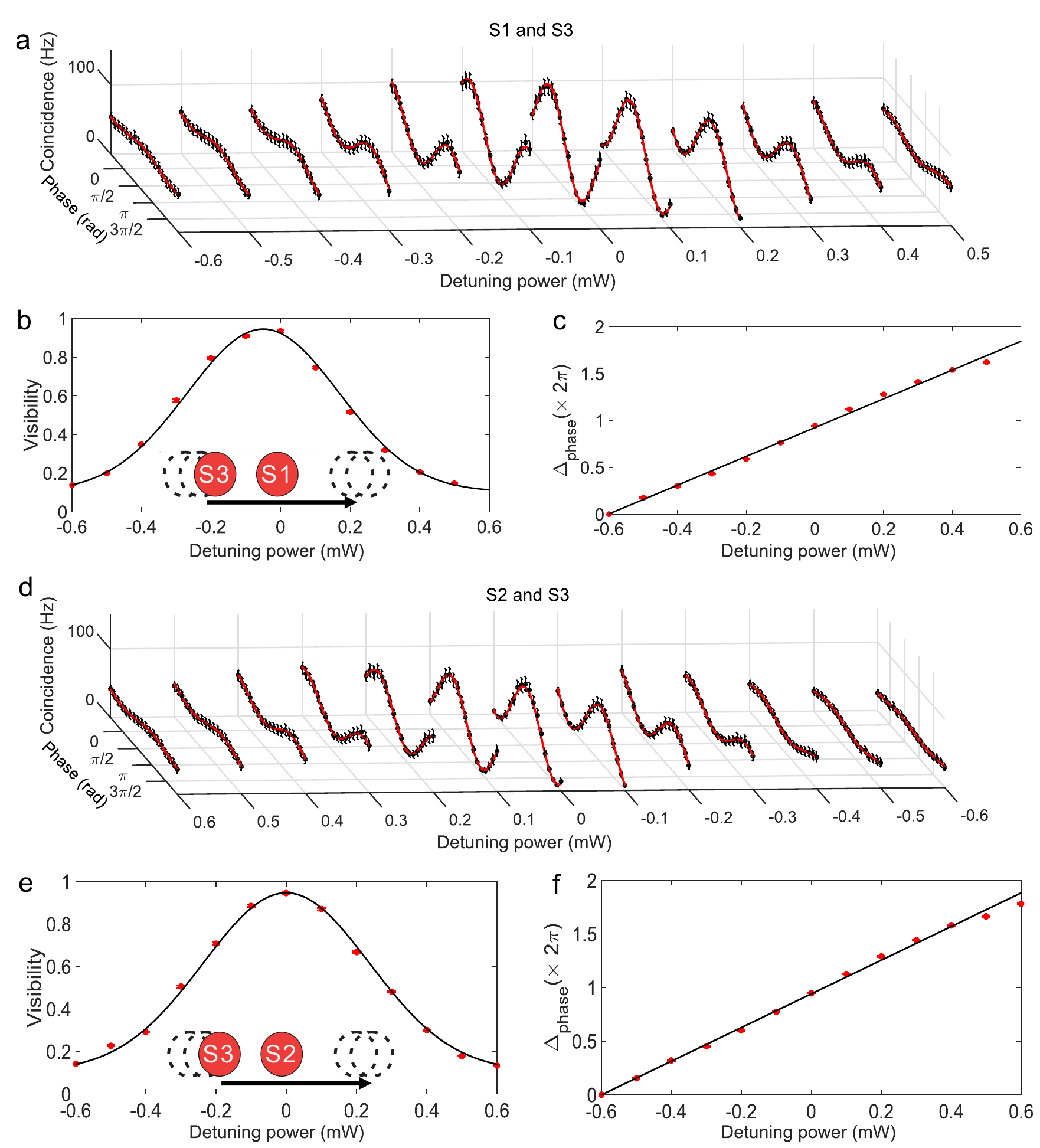}
	\caption*{Supplementary Figure 8. Path correlations results for S1 and S3 and S2 and S3 as one source resonance is swept over the other. Two-qubit correlation curves (\textbf{a/d}), changes in the two-qubit path interference visibilities (\textbf{b/e}) and phase variations between the different interference curves (\textbf{c/f}) are measured as S3 is scanned over S1/S2 with a step of 0.1 mW. The error bars are calculated by a Poissonian distribution. The red dots are the measured data, and the curves are drawn to guide the eye. }
\end{center}
\end{figure}

Supplementary Figure 8a to 8c show the two-qubit correlation curves with variations in the interference fringes, i.e., visibilities (b), and phase variations (c) between the different interference curves as a function of the S3-to-S1 detuning. Supplementary Figure 8d to 8f show the corresponding results of the S3-to-S2 detuning. To maintain the matching conditions between MZI1, MZI2 and the ring, we simultaneously change the three heaters on the scanned source with a step of 0.1 mW. It is clear that when the sources are largely detuned, no obvious interference occurs. As the spectra of the sources gradually match, clear interference patterns can be observed. In Supplementary Figure 8c and 8f, the phase variations between the different interference curves are almost linear with approximately the same dependence on the source scanning power, approximately $\frac{10 \pi}{3}mW^{-1}$. Therefore, if there is non-negligible thermal or electrical crosstalk mismatching the spectra of the sources, the initial phase of the interference curves will change.

\begin{figure}[!htbp]
	\includegraphics[width=0.8\textwidth]{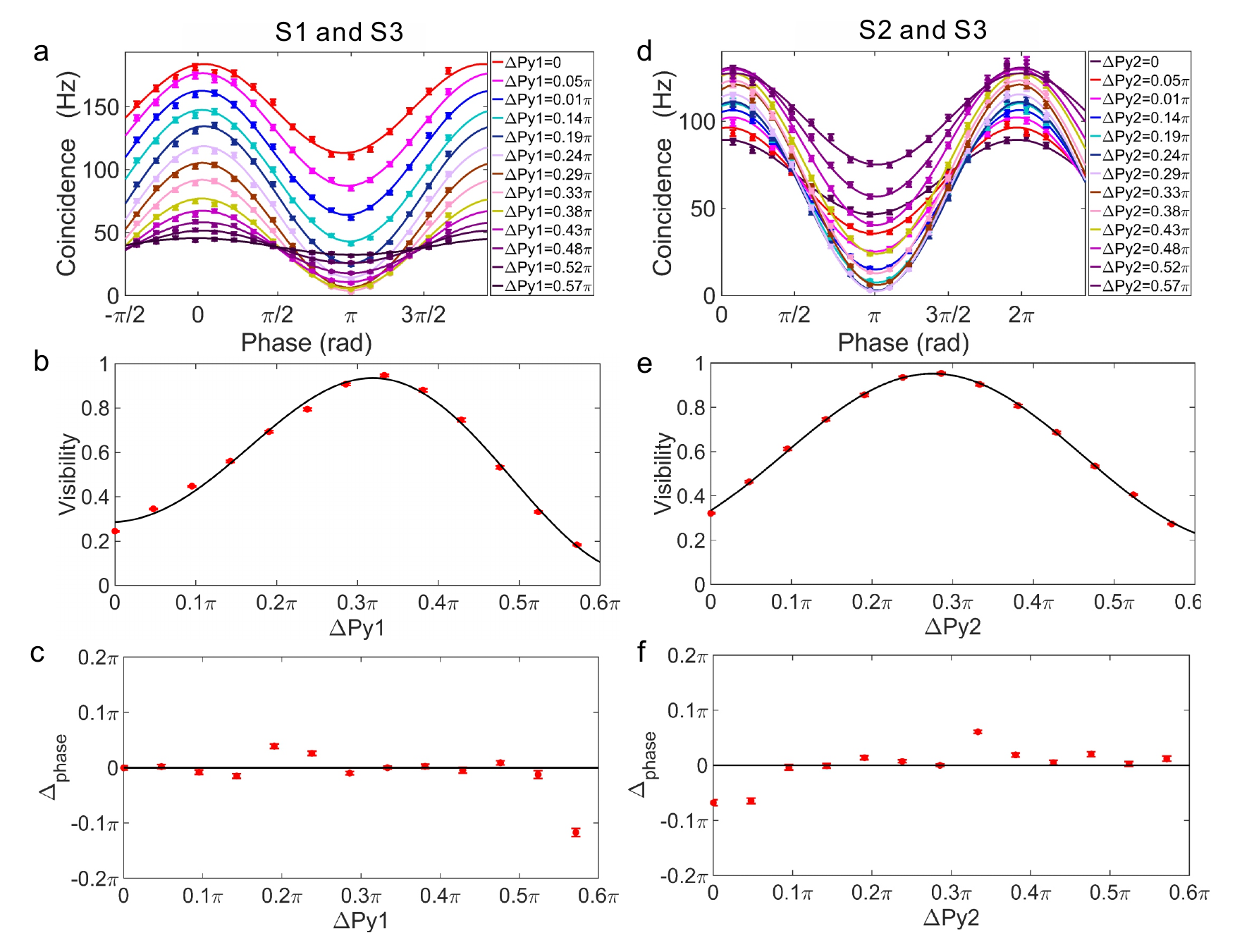}
	\caption*{Supplementary Figure 9. Path correlations for S1 and S3 (a-c) and S2 and S3 (d-f), obtained for different current conditions (with a 1 mW step)
   applied to $P_{y1}$ and $P_{y2}$, used as a pump power splitter to balance the generation rate between the measured sources.
   The interference fringes (\textbf{a,d}), visibilities (\textbf{b,e}) and changes in the initial phase between the different interference curves (\textbf{c,f})
   are shown for different pump power splitting ratios. The error bars are calculated by a Poissonian distribution.
   The points represent the experimental data, and the curves are fitted to guide the eye.}
\end{figure}

In Supplementary Figure 9, the fringes, visibilities and phase variations between the different path correlation curves for S1 and S3 (a-c) and S2 and S3 (d-f) are plotted for different pump power splittings between the sources while setting the sources detuning power to 0. The step of the heating power applied to $P_{y1}$ or $P_{y2}$ is 1 mW, and the phase change is approximately 0.048$\pi$. As can be seen from a and d, the interference curves exhibit a significant dependence on the balance of the photon-pair emission between the sources, which can be used to control the amplitude between the components of the output state to prepare variably entangled states. When a maximally entangled state is produced, the maximum visibility can be observed in b and e (94.72\%$\pm$0.50\% for S1 and S3, 96.10\%$\pm$0.40\% for S2 and S3, respectively). It is also worth noting that the phase variations shown in c and f nearly do not change as the balance changes, indicating that the scanned PS has little impact on the resonator-based sources, since changes in the spectra overlap, i.e., resonance shifts between the sources can introduce a relative phase between the two-qubit state, as shown in Supplementary Figure 8. This shows that the electrical and thermal crosstalk are effectively isolated.

\textbf{Two-qutrit correlation space}

Generally, for a two-qutrit system, the entangled state can be written as
\begin{equation}\label{eq18}
\Psi=(\alpha \exp^{i2\varphi_{pz1}}\left|00\right\rangle +\beta \exp^{i2\varphi_{pz2}}\left|11\right\rangle +\gamma \left|22\right\rangle ).
\end{equation}
A complete characterization of the higher-order two-qutrit correlations can be realized by using two three-dimensional multiports (3D-MPs). After each qutrit enters the 3D-MP, the state is transformed by a local unitary transformation ($\mathbb{U}$):
\begin{equation}\label{eq19}
\mathbb{U}=\frac{1}{3}\left(
\begin{array}{ccc}
1&1&1\\
1&e^{i\frac{2\pi}{3}}&e^{-i\frac{2\pi}{3}}\\
1&e^{-i\frac{2\pi}{3}}&e^{i\frac{2\pi}{3}}\\
\end{array}\right) \otimes
\left(
\begin{array}{ccc}
1&1&1\\
1&e^{i\frac{2\pi}{3}}&e^{-i\frac{2\pi}{3}}\\
1&e^{-i\frac{2\pi}{3}}&e^{i\frac{2\pi}{3}}\\
\end{array}\right)
\end{equation}

Then, the resulting state can be written as
\begin{equation}\label{eq20}
\begin{aligned}
\ket{\Psi^{'}} &=U\ket{\Psi} \\
&=\frac{1}{3}\alpha e^{i2\varphi_{pz1}}\left| \begin{array}{c}
1 \\
1 \\
1 \end{array}
\right\rangle \otimes\left| \begin{array}{c}
1 \\
1 \\
1 \end{array}
\right\rangle +\frac{1}{3}\beta e^{i2\varphi_{pz2}}\left| \begin{array}{c}
1 \\
e^{i2\pi/3} \\
e^{-i2\pi/3} \end{array}
\right\rangle \otimes\left| \begin{array}{c}
1 \\
e^{i2\pi/3} \\
e^{-i2\pi/3} \end{array}
\right\rangle +\frac{1}{3}\gamma\left| \begin{array}{c}
1 \\
e^{-i2\pi/3} \\
e^{-i2\pi/3} \end{array}
\right\rangle \otimes\left| \begin{array}{c}
1 \\
e^{-i2\pi/3} \\
e^{-i2\pi/3} \end{array}
\right\rangle
\end{aligned}
\end{equation}
The probabilities of detecting a photon pair for different detector combinations are:
\begin{equation}\label{eq21}
	\begin{split}
		P_{12}=P_{36}=P_{54}&=\frac{1}{9}{\left|\alpha e^{i(2\varphi_{pz1}+\frac{2\pi}{3})}+\beta e^{i(2\varphi_{pz2}-\frac{2\pi}{3})}+\gamma\right|}^2\\
		P_{14}=P_{32}=P_{56}&=\frac{1}{9}{\left|\alpha e^{i2\varphi_{pz1}}+\beta e^{i2\varphi_{pz2}}+\gamma \right|}^2 \\
        P_{16}=P_{34}=P_{52}&=\frac{1}{9}{\left|\alpha e^{i(2\varphi_{pz1}-\frac{2\pi}{3})}+\beta e^{i(2\varphi_{pz2}+\frac{2\pi}{3})}+\gamma\right|}^2.
	\end{split}
\end{equation}

\begin{figure}[!htbp]
	\includegraphics[width=0.8\textwidth]{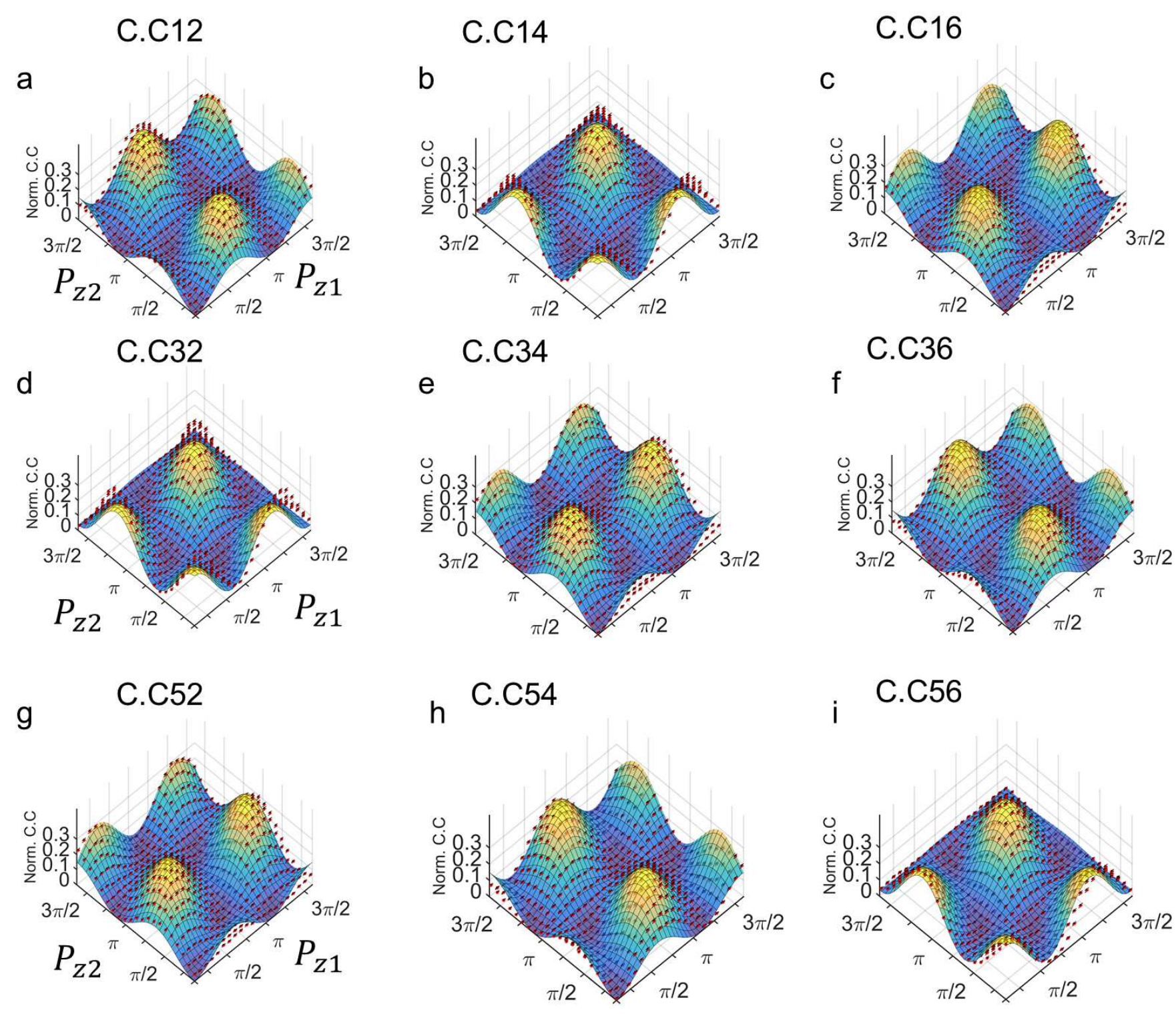}
	\caption*{Supplementary Figure 10. Full scan of the correlation space for two entangled qutrits state
   $\ket{\Psi}=\frac{1}{\sqrt{3}}(e^{i2\varphi_{pz1}}\left|00\right\rangle +e^{i2\varphi_{pz2}}\left|11\right\rangle +\left|22\right\rangle)$.
   Figures \textbf{a} to \textbf{i} correspond to detector combinations of (1,2), (1,4), (1,6), (3,2), (3,4), (3,6), (5,2), (5,4), and (5,6) respectively,
   while two relative pump phases ${\varphi }_{pz1}$ and
   ${\varphi }_{pz2}$ are scanned. The error bars are calculated by a Poissonian distribution. The normalized measured coincidences
   (Norm. C.C) (red dots) and simulated results (lines) are compared and show excellent agreement with each other.}
\end{figure}

In the experiment, we set $\alpha=\beta=\gamma=\frac{1}{\sqrt{3}}$ and adjust the relative pump phase over S1 and S2 to characterize the high-order Einstein-Podolsky-Rosen (EPR) correlations. All nine output combinations of the coincidences and their respective theoretical results are shown in Supplementary Figure 10. Due to the symmetry of the multi-mode interferometers, three probabilities $P_{ij}$ of equal values are observed. For each pattern, there are three repeated results for each correlation pattern, and the experimental results are in good agreement with the theoretical results.

\section*{Supplementary References}

\end{document}